 \documentclass[a4paper,twocolumn,11pt,accepted=2025-10-08]{quantumarticle}
 \pdfoutput=1
\usepackage[utf8]{inputenc}
\usepackage{graphicx}
\usepackage{float}
\usepackage{dcolumn}
\usepackage{bm}
\usepackage{physics}
\usepackage{dsfont}
\usepackage{todonotes}

\usepackage[colorlinks,linkcolor=blue,urlcolor=blue, citecolor=blue]{hyperref}
\usepackage{xurl}
\usepackage[whole]{bxcjkjatype}
\usepackage{mathtools}
\usepackage{amsmath,amssymb}
\usepackage{comment}

\newcommand{\mF}{\mathcal{F}}

\newcommand{\bal}{\begin{equation}\begin{aligned}}
\newcommand{\eal}{\end{aligned}\end{equation}}
\newcommand{\sbar}{\;\rule{0pt}{9.5pt}\right|\;}
\newcommand{\lset}{\left\{\left.}
\newcommand{\rset}{\right\}}
\usepackage[numbers,sort&compress]{natbib}

\DeclareMathOperator{\cl}{cl}
\DeclareMathOperator{\conv}{conv}

\begin{document}

\title{Classical simulation and quantum resource theory of non-Gaussian optics}

\author{Oliver Hahn}
\email{hahn@g.ecc.u-tokyo.ac.jp}
\orcid{0000-0003-1677-8696}
\affiliation{Wallenberg Centre for Quantum Technology, Department of Microtechnology and Nanoscience, Chalmers University of Technology, Sweden , SE-412 96 G\"{o}teborg, Sweden}
\affiliation{Department of Basic Science, The University of Tokyo, 3-8-1 Komaba, Meguro-ku, Tokyo, 153-8902, Japan}
\author{Ryuji Takagi}
\orcid{0000-0003-3837-8159}
\affiliation{Department of Basic Science, The University of Tokyo, 3-8-1 Komaba, Meguro-ku, Tokyo, 153-8902, Japan}

\author{Giulia Ferrini}
\orcid{0000-0002-7130-6723}
\affiliation{Wallenberg Centre for Quantum Technology, Department of Microtechnology and Nanoscience, Chalmers University of Technology, Sweden , SE-412 96 G\"{o}teborg, Sweden}

\author{Hayata Yamasaki}
\affiliation{Department of Physics, Graduate School of Science,
The University of Tokyo, 7–3–1 Hongo, Bunkyo-ku, Tokyo, 113–0033, Japan}
\orcid{0000-0003-3521-831X}
\affiliation{Department of Computer Science, Graduate School of Information Science and Technology, The University of Tokyo, 7-3-1 Hongo, Bunkyo-ku, Tokyo 113-8656, Japan}

\begin{abstract}
We propose efficient algorithms for classically simulating Gaussian unitaries and measurements applied to non-Gaussian initial states. 
The constructions are based on decomposing the non-Gaussian states into linear combinations of Gaussian states.
We use an extension of the covariance matrix formalism to efficiently track relative phases in the superpositions of Gaussian states. We get an exact simulation algorithm, which costs quadratically with the number of Gaussian states required to represent the initial state, and an approximate simulation algorithm, which costs linearly with the $l_1$ norm of the coefficients associated with the superposition. We define measures of non-Gaussianity quantifying this simulation cost, which we call the Gaussian rank and the Gaussian extent. From the perspective of quantum resource theories, we investigate the properties of this type of non-Gaussianity measure and compute optimal decompositions for states relevant to continuous-variable quantum computing.
\end{abstract}

\maketitle

\section{Introduction}
\label{sec:introduction}

Quantum computing holds the promise of solving problems that are currently beyond the reach of classical methods. 
In recent years, continuous-variable (CV) systems have emerged as a prime candidate to implement quantum computation and have received increased attention due to their inherent noise resilience, showcased by a series of breakthrough experiments~\cite{Ofek:2016wb,sivak2023real}. CV systems cannot
be described within a finite-dimensional Hilbert space and naturally appear in optical~\cite{pfister2019continuous}, microwave radiation~\cite{blais2020circuit, grimsmo2017squeezing}, and optomechanical systems~\cite{schmidt2012optomechanical, houhou2015generation}.

In practical implementations, some operations are easier to perform experimentally than others.
While Gaussian operations are relatively straightforward to implement in CV systems, they are not sufficient for achieving universal quantum computing. Moreover, when restricted to Gaussian operations and states, efficient classical simulation becomes possible~\cite{mari2012positive} and tasks such as error correction become impossible~\cite{PhysRevLett.102.120501}. Thus, the incorporation of non-Gaussian resources, such as Fock states, becomes necessary for achieving universal quantum computing and opens the possibility for any potential quantum advantage.

 This fact introduces a big challenge in assessing and designing new quantum information protocols, such as error correction schemes. Indeed, quantum mechanical systems are notoriously difficult to classically simulate in general, even more so in the case of CV systems.
Aside from such practical use cases, classical simulation algorithms are useful in tackling fundamental problems.
 They provide a lens to directly study the boundary between the computational power of classical and quantum computational models and help to identify the origin of potential quantum advantages.

The other side of the coin is that adding non-Gaussian resources to otherwise classically simulatable Gaussian systems in turn makes it hard to simulate them classically. Thus, any classical simulation algorithm will be restricted in some way. One way to simulate a CV system is to use the Wigner function~\cite{Veitch_2013,Pashayan2015estimating}. The simulation time scales with the amount of negative parts of the Wigner function.
The negativity of the Wigner function~\cite{Kenfack2004negativity,takagi2018convex, albarelli2018resource} is a measure of non-Gaussianity, meaning that Gaussian states and processes can be represented with strictly positive Wigner function and thus sampled from efficiently. Hence, this method is intrinsically constrained to circuits with very low or no Wigner negativity. The simulation of specific circuits with large negativity has been considered in Refs.~\cite{garcia-alvarez2020efficient, calcluth2022efficient, calcluth2023vacuum,PRXQuantum.6.010330}. The approaches therein are restricted to the simulation of a Gaussian circuit with input infinitely squeezed GKP ``states'' encoding stabilizer states, and are not directly applicable to the case of more general input non-Gaussian states, including realistic (finitely-squeezed) GKP states.  
A further approach relies on decomposing states in the Fock basis with bounded support~\cite{PhysRevResearch.3.033018, PhysRevLett.130.090602}.
The non-Gaussian measure quantifying the simulation overhead of non-Gaussian states using such a decomposition is the stellar rank~\cite{PhysRevLett.124.063605}.
However, many states that are of utmost relevance for CV quantum computing, such as GKP and cat states, require a large number or even infinitely many Fock states, leading to scaling issues. The simulator~\cite{bourassa2021fast} circumvents some of these issues, but its performance is not studied analytically, making an analysis difficult.

In this work, we introduce two classical algorithms for the simulation of CV systems, which are inherently connected with two measures for the resource theory of non-Gaussianity.
The idea behind the simulators is to decompose non-Gaussian states into a superposition of Gaussian states and use an efficient Gaussian subroutine. For the subroutine, we develop a technique to compute the relative phase of these Gaussian states using an extension of the covariance matrix formalism.
 The first algorithm is exact and scales with the number of Gaussian states in the superposition---the Gaussian rank---quadratically.
 We improve this algorithm by approximating the input state by sparsifying the decomposition, yielding the second algorithm. For the latter, we also develop a method for the fast estimation of the norm of the non-Gaussian state, bringing the scaling down to linear in the $l_1$ norm of the coefficients in the superposition, or equivalently, the Gaussian extent.
 Hence, in both algorithms, the simulation cost increases with the amount of non-Gaussianity in the input state, measured by the two measures of non-Gaussianity we introduce.

 Our simulation algorithms are particularly suitable for quantum optical setups, where non-Gaussian states are generated and then transformed via Gaussian operations.
Although we consider primarily the scenario of input non-Gaussian states followed by Gaussian operations and measurements, the simulators also allow for the simulation of other non-Gaussian circuits through gadgetisation.
 
The idea of the simulator, through decomposing non-Gaussian states into superpositions of Gaussian states, is based on analogous ideas that were used for the simulation of magic states in qubit systems~\cite{bravyi2019simulation},  fermionic~\cite{dias2023classical}, and passive linear optics~\cite{Marshall2023simulation}.

In the second part, we formally introduce the two measures, the Gaussian rank and the Gaussian extent. These measures quantify the smallest cost to run the two algorithms.
This fact gives the Gaussian rank and the Gaussian extent an immediate operational interpretation.
Additionally, we thoroughly investigate the properties of the measures. The Gaussian extent is connected to the generalized robustness of non-Gaussianity~\cite {Regula2021operational,Lami2021framework}. 
This allows us to derive a useful condition for the optimal witness of the generalized robustness.
At last, we compute optimal decompositions of states of interest for bosonic error correction, as well as derive bounds for CV protocols such as Gaussian boson sampling and cat state breeding.
Our work connects the practical usefulness of an efficient classical simulator with the fundamental investigation of resources required for CV quantum computing.

The rest of this paper is organized as follows.
In Sec.~\ref{sec:background}, we provide an introduction to Gaussian quantum optics.
In Sec.~\ref{sec:phase}, we develop a phase-sensitive simulator for Gaussian states and operations based on the covariant matrix formalism. This allows us to compute the overlap between pure Gaussian states, including the phase information, which is not available using the standard formalism based on the covariance matrix.
In Sec.~\ref{sec:simulatingalg}, we provide two classical simulation algorithms for non-Gaussian optics. The first algorithm performs exact simulation by using the decomposition of non-Gaussian states into a superposition of Gaussian states. This allows the usage of continuous decompositions as well.
The second algorithm improves the runtime scaling by using a low-rank approximation of non-Gaussian states by sparsifying their decomposition in Gaussian states. These low-rank approximations are generally not normalized, so we introduce a fast norm estimation algorithm. 
In Sec.~\ref{sec:measures} we introduce two measures in the resource theory of non-Gaussianity --- the Gaussian rank and the Gaussian extent --- and prove their properties. The simulation cost of the two simulation algorithms we introduced scales with the resource measured by the two new measures. We characterize the Gaussian extent by a robustness measure and find a condition that needs to be obeyed by the optimal decomposition in Gaussian states.
Examples and applications of this measure are given in Sec.~\ref{sec:applications}, such as optimal decompositions and a resource-theoretic analysis using the Gaussian extent of Gaussian boson sampling and cat state breeding.
Finally, our conclusions are given in Sec.~\ref{sec:conclusion}.

\section{Gaussian quantum optics}
\label{sec:background}
Gaussian quantum optics offers a rich area of research.
Beyond its vast range of applicability, which spans from quantum metrology to quantum key distribution, an appealing feature of Gaussian quantum optics is that several analytical techniques are available in this regime. Here, we introduce the main notations and formalism used in this paper regarding Gaussian quantum optics.

In this work, we will use the canonical operators $q,p$ with
\begin{align}
    [q,p]=i.
\end{align}

Gaussian unitaries are defined as 
\begin{align}
    U&=e^{i H}\\
    H&= \frac{1}{2}\bm{r}^T\text{H}\bm{r}+\bm{\Bar{r}}\bm{r}
\end{align}
where we use the shorthand notation $\bm{r}=(q_1,p_1,...,q_n,p_n)^T$ for the vector of canonical operators and $\bm{\Bar{r}}= (r_{q_1},r_{p_1},...r_{q_n},r_{p_n})^T$ a vector of real numbers.
The matrix $\text{H}$ is a symmetric matrix, while $H$ denotes the operator.

This allows us to define Gaussian states as 
\begin{align}
    \rho_G=\frac{e^{-\beta H}}{\Tr[e^{-\beta H}]}
\end{align}
including the case $\beta\rightarrow\infty$ being pure states.
We let $\mathcal{G}$ denote the set of Gaussian states.

The advantage of Gaussian states is that they are fully determined by the mean
\begin{align}
    \bm{\Bar{r}}=\Tr[\bm{r}\rho]
\end{align}
and the covariance matrix
\begin{align}
    \sigma= \Tr[\{(\bm{r}-\bm{\Bar{r}}),(\bm{r}-\bm{\Bar{r}})^T\}\rho]
\end{align}
where $\{,\}$ is the anti-commutator, and $\bm{r} \bm{r}^T$ is the outer product.
The requirement on all covariance matrices is
\begin{align}
    \sigma\pm i\Omega\geq0,   
\end{align}
where $\Omega = \bigoplus^{N}_{j=1} \begin{pmatrix}
    0& 1\\
    -1 &0 
    \end{pmatrix}$.
Gaussian operations refer to any operations composed of the preparation of Gaussian states, applications of Gaussian unitaries, and measurement by homodyne or heterodyne detection.

Using the covariance matrix formalism, one can easily compute the dynamics of a quantum state as long as the operations involved are Gaussian.
This fact means that Gaussian quantum optics can be efficiently simulated on a classical computer.
This implies, in turn, that no exponential quantum advantage can be achieved with only Gaussian quantum circuitry.
Below, we summarize a list of ingredients used for the Gaussian simulator in the covariance matrix formalism~\cite{serafini2017quantum}.

\paragraph{Gaussian unitaries}
Displacement operations by $\bm{\Bar{r}^\prime}$ have the following action on the mean and the covariance matrix:
\begin{align}
    \bm{\Bar{r}}&\rightarrow \bm{\Bar{r}}+\bm{\Bar{r}^\prime},\\
    \sigma&\rightarrow\sigma.
\end{align}
A general symplectic transformation $S$ acts as
\begin{align}
   \bm{ \Bar{r}}&\rightarrow S\bm{ \Bar{r}},\\
    \sigma&\rightarrow S\sigma S^T.
\end{align}
There exists a parameterization for symplectic transformations~\cite{dopico2009parametrization}.
Every symplectic transformation can be decomposed as~\cite{seddon_quantifying_2021}
\begin{align}
    S&=O_1 ZO_2,\\
    Z&=\bigoplus_{i=1}^n \begin{pmatrix}
        z_i&0\\
        0& z_i^{-1}
    \end{pmatrix},
\end{align}
with $O_1,O_2 \in SP_{2n,\mathds{R}}\cap SO(2n)$, which are the passive symplectic transformations.
So we can decompose all symplectic transformations in terms of single-mode squeezers, phase shifters, and two-mode beam splitters.
Note that $ SP_{2n,\mathds{R}}\cap SO(2n)$  is isomorphic to $U(n)$
\begin{align}
    O_i = \Bar{U}'^{\dagger}   \begin{pmatrix}
        U^*&0_n\\
        0_n& U
    \end{pmatrix}
    \Bar{U}'
\end{align}
with $U\in U(n)$ and
\begin{align}
    \Bar{U}'=\frac{1}{\sqrt{2}}\begin{pmatrix}
        \mathds{1}_n& i\mathds{1}_n\\
        \mathds{1}_n & -i\mathds{1}_n
    \end{pmatrix}.
\end{align}

\paragraph{Tensor product and partial trace}
Suppose that we have a two-mode Gaussian state with
\begin{align}
    \bm{ \Bar{r}}&= \bm{ \Bar{r}_A} \oplus\bm{ \Bar{r}_B}=\begin{pmatrix}
       \bm{ \Bar{r}_A}\\
        \bm{ \Bar{r}_B}
    \end{pmatrix},\\
    \sigma &= \sigma_A \oplus \sigma_B =\begin{pmatrix}
        \sigma_A&0\\
        0&\sigma_B
    \end{pmatrix}.
\end{align}
Then, the partial trace on the second subsystem yields
\begin{align}
    \bm{ \Bar{r}}& \rightarrow \bm{ \Bar{r}_A},\\
    \sigma& \rightarrow \sigma_A.
\end{align}

\paragraph{Gaussian completely positive and trace-preserving (CPTP) map}
Given a Gaussian initial state with covariance $\sigma$ and mean $\bm{ \Bar{r}}$, the evolution of a Gaussian CPTP map is characterized by two real matrices $X,Y$ and a vector $\bm{D}$ as
\begin{align}
    \bm{ \Bar{r}}&\rightarrow X\bm{ \Bar{r}}+\bm{D},\\
    \sigma&\rightarrow X\sigma X^T +Y,
\end{align}
with the requirement that
\begin{align}
    Y+i\Omega\geq i X\Omega X^T.
\end{align}

\paragraph{Gaussian measurements}
Measurement results of homodyne detection are determined by the mean and covariance matrix, e.g.,
\begin{align}
    p(x)= \frac{e^{-\frac{(x-\Bar{x})^2}{\sigma_1}}}{\sqrt{\pi \sigma_1}}.
\end{align}
More generally, the POVMs of a general-dyne detection are given with the completeness condition
\begin{align}
    \mathds{1}_n= \frac{1}{(2\pi)^{2n}}\int_{\mathds{R}^{2n}} \dd\bm{r_m} D(-\bm{r_m}) S^\dagger \dyad{0}SD(\bm{r_m})
\end{align}
with the outcomes $\bm{r_m}$.
The probability density $p(\bm{r_m})$ of measuring a Gaussian state $\rho$ with $\sigma$ and $\bm{\Bar{r}}$ can then be inferred from the expression of Gaussian overlaps
\begin{align}
    p(\bm{r_m}) = \frac{\ev{\rho}{\psi_G}}{(2\pi)^n}= \frac{e^{-(\bm{r_m}-\bm{\Bar{r}})^T (\sigma+SS^T)^{-1}   (\bm{r_m}-\bm{\Bar{r}})}}{\pi^n  \sqrt{\det\qty(  \sigma +SS^T ) } },
\end{align}
where $\ket{\psi_G} =D(-\bm{r_m}) S^\dagger \ket{0} $ from the POVM\@.
One retrieves homodyne detection in certain limits; e.g, for $n=1$, one measures $q$ in the limit $SS^T = \lim_{z\rightarrow0} \text{diag} (z^2, 1/z^2)$ and $p$ for $z\rightarrow \infty$.

One can generalize this notion to noisy measurements, such as measurements with finite detection efficiency. The POVMs are then given with the completeness condition
\begin{align}
    \mathds{1}_n= \frac{1}{(2\pi)^{2n}}\int_{\mathds{R}^{2n}} \dd\bm{r_m}D(-\bm{r_m}) \rho_m D(\bm{r_m})
\end{align}
for the outcomes $\bm{r_m}$ and $\rho_m$ having covariance matrix $\sigma_m$.
The probability density of obtaining result $\bm{r_m}$ is then
\begin{align}
    p(\bm{r_m})=\frac{1}{(2\pi)^n}\Tr \qty[  \rho D(-\bm{r_m})  \rho_m D(\bm{r_m}) ]
\end{align}
or equivalently if $\rho$ is a Gaussian state with covariance $\sigma$ and mean $\bm{\Bar{r}}$
\begin{align}
    p(\bm{r_m})= \frac{e^{-(\bm{r_m}-\bm{\Bar{r}})^T  (\sigma+\sigma_m)^{-1} (\bm{r_m}-\bm{\Bar{r}})}}{\pi^n \sqrt{\det\qty[\sigma+\sigma_m]}}.
\end{align}

\paragraph{Conditional Gaussian dynamics}
We consider a bipartite Gaussian state
\begin{align}
\bm{\Bar{r}}&=\begin{pmatrix}
    \bm{\Bar{r}}_A\\
    \bm{\Bar{r}}_B
\end{pmatrix},\\
    \sigma&=\begin{pmatrix}
        \sigma_A &\sigma_{AB}\\
        \sigma_{AB}^T& \sigma_B
    \end{pmatrix}.
\end{align}
The $n$-mode subsystem of an $(n+m)$-mode Gaussian state with the covariance and mean defined above undergoes the following mapping upon general-dyne measurement of the $m$ modes characterized by measurement outcome $\bm{r_m}$ and covariance $\sigma_m$:
\begin{align}
    \bm{\Bar{r}}_A &\rightarrow\bm{\Bar{r}}_A -\sigma_{AB} \frac{1}{\sigma_B+\sigma_m}(\bm{r_m}-\bm{\Bar{r}}_B),\\
    \sigma_A &\rightarrow \sigma_A-\sigma_{AB}\frac{1}{\sigma_B+\sigma_m}\sigma_{AB}^T.
\end{align}

\section{Phase-sensitive simulator for Gaussian states and operations}
\label{sec:phase}
First, we consider the simulation when all states and operations are Gaussians.
The task is to compute the Born probability of obtaining a sample $x$ while performing heterodyne detection of the Gaussian state $\ket{G}$
\begin{align}
    P(x)=\abs{\braket{G}{x}}^2.
\end{align}
In principle, this can easily be computed using the covariance formalism, as it is well known that Gaussian states and operations are classically simulable efficiently.

However, the issue is that the conventional way of simulating Gaussians as summarized in Sec.~\ref{sec:background} is based on the covariance matrix formalism and thus does not compute global phases.
By contrast, for our non-Gaussian simulator, as we will see later, we need to exploit a Gaussian simulator that keeps track of the phases, which we will develop here inspired by Ref.~\cite{Bravyi2017complexity}. A more in-depth derivation can be found in Appendix~\ref{ap:innerproduct}.

The most important ingredient is to compute the inner product of Gaussian pure states
\begin{align}
    \bra{G_i}\ket{G_j}\quad\text{for $\ket{G_i},\ket{G_j}\in \mathcal{G}$},
\end{align}
in a phase-sensitive way; i.e., our method does not only provide $|\bra{G_i}\ket{G_j}|$ but also clarifies $\theta\in[0,2\pi)$ for $\bra{G_i}\ket{G_j}=e^{i\theta}|\bra{G_i}\ket{G_j}|$.
We fix a Gaussian reference state $G_0 = \dyad{G_0}$, which can be an arbitrary Gaussian state.

Given the reference state $G_0 = \dyad{G_0}$ and our two Gaussian states of interest $G_1=\dyad{G_1}$ and $G_2=\dyad{G_2}$, we deduce the phase sensitive overlap between $\bra{G_1}\ket{G_2}$ via
\begin{align}
    \Tr \qty(G_0 G_1 G_2)&=\Tr\qty(\dyad{G_0}\dyad{G_1}\dyad{G_2})\\
    \label{eq:G_0G_1G_2}
    &=\bra{G_2}\ket{G_0}\bra{G_1}\ket{G_2}\bra{G_0}\ket{G_1}.
\end{align}
The fidelity between an $n$-mode Gaussian state $\rho_0$ with covariance $\sigma_0$ and mean $\bm{x_0}$ and an $n$-mode pure Gaussian state $\rho_1=\dyad{\phi}$ with covariance $\sigma_1$ and mean $\bm{x_1}$ is given by~\cite{spedalieri2012limit}
\begin{align}
    \mathcal{F}(\rho_0,\dyad{\phi}) = \bra{\phi}\rho_0\ket{\phi} = 2^n \frac{e^{-\qty(\bm{d}^T(\sigma_0+\sigma_1)^{-1}\bm{d} )}}{\sqrt{\det\qty[\sigma_0+\sigma_1]}}
\end{align}
with $\bm{d}= \bm{x_0}-\bm{x_1}$. 
Then, as shown in Appendix~\ref{ap:innerproduct}, the right-hand side of~\eqref{eq:G_0G_1G_2} can be calculated using the covariance matrices $\sigma_i$ and means $\bm{\mu_i}$ by
\begin{widetext}
\begin{align}
    \bra{G_2}\ket{G_0}\bra{G_1}\ket{G_2}\bra{G_0}\ket{G_1} &= \frac{1}{\sqrt{\det(\sigma_1+\sigma_2)}} \frac{1}{\sqrt{\det(\sigma_0+\Delta)}} e^{-\frac{1}{4}(\bm{\mu_1}-\bm{\mu_2})^T(\sigma_1+\sigma_2)^{-1}  (\bm{\mu_1}-\bm{\mu_2})}  \\
    &e^{-\frac{1}{4}(\bm{\mu_0}-\mu_{\Delta})^T(\sigma_0+\Delta)^{-1}(\bm{\mu_0}-\mu_{\Delta}) },
\end{align}
where we write
\begin{align}
    \Delta&=\sigma_2-  \frac{(\sigma_2-i\Omega)^T}{2}(\sigma_1+\sigma_2)^{-1} \frac{(\sigma_2-i\Omega)}{2},\\
    \label{eq:mu}
    \bm{\mu_{\Delta}}&=\bm{\mu_2}+\frac{1}{2}(\bm{\mu_1}-\bm{\mu_2}) (\sigma_1+\sigma_2)^{-1}  \frac{(\sigma_2-i\Omega)}{2} .
\end{align}

\end{widetext}
In order to obtain the desired inner product $\bra{G_1}\ket{G_2}$, we need to use the inner products with the reference states.
Therefore, we not only specify all Gaussian states with covariance matrix $\sigma_i$ and mean $\bm{\mu_i}$, but we also use the inner product with the reference state, which we indicate as $o_i$; as a whole, we will henceforth represent a pure Gaussian state as
\begin{equation}
\label{eq:G_i}
   \ket{G_i}=\ket{\sigma_i,\bm{\mu_i},o_i} 
\end{equation}
when necessary. 
We can then use the tools presented in Sec.~\ref{sec:background} to update the covariance and mean of the Gaussian state and then perform a Gaussian measurement by computing the overlap with the corresponding Gaussian state in a phase-sensitive manner.
We can directly give the covariance matrices and means to compute the overlap with the reference state for a single mode; more generally, we give an algorithm to do this computation in Appendix~\ref{ap:refstate}.
Also in Appendix~\ref{ap:alt_phase}, we give based on the article~\cite{yao2024riemannianoptimizationphotonicquantum} an alternative way to simulate Gaussian overlaps in a phase sensitive way using the stellar formalism.

\section{Classical simulation algorithm for non-Gaussian optics}
\label{sec:simulatingalg}
In this section, we present our non-Gaussian simulation algorithm using the phase-sensitive Gaussian simulator introduced in Sec.~\ref{sec:phase}.
Our interest is the classical simulation of representative classes of operations in non-Gaussian optics that can be written as combinations of Gaussian operations with auxiliary non-Gaussian states, e.g., quantum computation using Gottesman-Kitaev-Preskill (GKP) codes~\cite{PhysRevA.64.012310,PhysRevLett.123.200502,PhysRevResearch.2.023270} and cubic phase gates implemented by gate teleportation~\cite{PhysRevA.64.012310,doi:10.1080/09500340601101575}.

As a simple case, we begin by considering non-Gaussian states decomposed into a sum of Gaussian states as
\begin{align}
\label{eq:decomposition}
    \ket{\psi}=\sum_{i=1}^\chi c_i \ket{G_i},
\end{align}
where we used the notation $\ket{G_i}=\ket{\sigma_i, \bm{\mu_i}, o_i}$ as in Eq.~\eqref{eq:G_i}.
The Born probability that we want to estimate then has the form
\begin{align}
\label{eq:born}
    P(x)=\frac{\braket{\psi}{x}\braket{x}{\psi} }{\norm{\psi}^2}= \frac{\sum_{i,j=1}^{\chi} c_j^* c_i\braket{G_j}{x}\braket{x}{G_i} }{\norm{\psi}^2},
\end{align}
where we perform heterodyne detection and $\ket{x}$ is then just a tensor product of coherent states.
We will then argue how to approximate arbitrary non-Gaussian states by such a finite sum.

\subsection{Exact simulation}

The first simulation algorithm computes the Born probability exactly by computing all overlaps in Eq.~\eqref{eq:born}.
Thus, to simulate Gaussian operations on non-Gaussian states, we update the individual Gaussian states using the tools of Sec.~\ref{sec:background}. Then we apply the phase-sensitive Gaussian simulator and the tools of Sec.~\ref{sec:phase} for each term $\bra{x}\ket{G_i}$ in Eq.~\eqref{eq:born},
where we need to take into account the relative phase between the terms.
The cost of exactly estimating the Born probability requires $\mathcal{O}(\chi^2)$ inner products.

\subsection{Approximate simulation}
\label{sec:fastnorm}
We can improve this simulation method by introducing an approximate version of the latter, by considering sampling terms $\ket{\sigma_i, \mu_i, o_i}$ in the decomposition Eq.~\eqref{eq:decomposition} to sparsify it.
The intuition behind this is to find low rank approximations and therefore speed up the computation. 
Using a sparsified state allows one to use continuous decompositions of the form
\begin{align}
    \ket{\psi} = \int{dk}\, c(k)\ket{\sigma_k,\bm{\mu_k},o_k}  
\end{align}
as well.
The sampling procedure becomes apparent if we rewrite the decomposition of the state $\psi$ as follows 
\begin{align}
\ket{\psi}&=\sum_{i=1}^\chi c_i\ket{\sigma_i, \mu_i,o_i}\\
        &=\norm{c}_1\sum_{i=1}^\chi p(i) \ket{\tilde{\sigma}_i, \tilde{\mu}_i, \tilde{o}_i},
\end{align}
where $\ket{\tilde{\sigma}_i, \bm{\tilde{\mu}_i}, \tilde{o}_i}\coloneqq(c_i/|c_i|)\ket{\sigma_i, \bm{\mu_i},o_i}$ with the 
probability distribution 
\begin{align}
p(i) &\coloneqq\frac{\abs{c_i}}{\norm{c}_1}.
\end{align}
The same holds for the continuous decomposition as well, with
\begin{align}
    \ket{\psi} =\norm{c}_1 \int{dk}\, p(k)\ket{\tilde{\sigma}_k, \bm{\tilde{\mu}_k}, \tilde{o}_k}
\end{align}
with the probability distribution
\begin{align}
p(k) &\coloneqq\frac{\abs{c(k)}}{\norm{c}_1}.
\end{align}

Thus if we sample Gaussian state  $\ket{\tilde{\sigma}_i, \tilde{\mu}_i, \tilde{o}_i}$ from the decomposition of $\ket{\psi}$ with probability $p(i)$, then the expectation is the state $\ket{\psi}$
\begin{align}
    \ket{\psi}=\norm{c}_1 \mathds{E}_{i\sim p(i)}\qty[\ket{\tilde{\sigma}_i, \bm{\tilde{\mu}_i}, \tilde{o}_i}],
\end{align}
where we may omit the subscript of the expectation value $\mathds{E}_{i\sim p(i)}$ to write $\mathds{E}$ if it is obvious from the context.

Sampling from this probability distribution $k$ times, we get a sparsified state---a low rank approximation of state $\ket{\psi}$---
\begin{align}
\label{eq:sparfied}
    \ket{\Omega}&=\frac{\norm{c}_1}{k}\sum_{i=1}^k\ket{\tilde{\sigma}_i, \bm{\tilde{\mu}_i}, \tilde{o}_i}.
\end{align}

As a consequence, we have that
\begin{align}
    \mathds{E}\qty[\bra{\Omega} \ket{\psi}] = \mathds{E}[  \bra{\psi }\ket{\Omega}] =1.
\end{align}

If we use the sparsified state $\ket{\Omega}$ instead of the state $\ket{\psi}$ in the computation the error is upper bounded by
\begin{align}
    &\mathds{E}[\abs{\ket{\psi}-\ket{\Omega}}^2 ]\nonumber\\
    &= \mathds{E}[\bra{\Omega}\ket{\Omega}] +\mathds{E}[\bra{\psi}\ket{\psi}]-\mathds{E}[\bra{\psi}\ket{\Omega}]-\mathds{E}[\bra{\Omega}\ket{\psi}]\\
    &\leq  \frac{\norm{c}_1^2}{k}.
\end{align}
Thus, by choosing the number
\begin{equation}
\label{eq:k}
    k=\qty(\frac{\norm{c}_1}{\delta})^2
\end{equation}
of sampled Gaussian states, we can upper-bound
\begin{equation}
    \mathds{E}\qty[\|\ket{\psi}-\ket{\Omega}\|^2 ] \leq \delta^2. 
    \label{eq:sparcification}
\end{equation}
The minimum sampling cost for a given state $\ket{\psi}$ and error $\delta$ in Eq.~\eqref{eq:k} for each non-Gaussian state is characterized by the Gaussian extent in Eq.~\eqref{eq:extent}.
Using the sparsified state in estimating the Born probability scales linearly with $\chi$ instead of quadratically in the exact case~\cite{bravyi2016improved}.
For further details and an alternative sampling strategy introduced in Ref.~\cite{seddon_quantifying_2021}, please consult  Appendix~\ref{ap:Gextent}.

However, in order to compute the Born probability in Eq.~\eqref{eq:born} approximately by using a sparsified state $\ket{\Omega}$ defined in Eq.~\eqref{eq:sparfied}, we need to compute the norm of the state. This is necessary because the sparsification will lead to an unnormalized state $\ket{\Omega}$.
Given a decomposition~\eqref{eq:sparfied} of $\ket{\Omega}$, the required time steps for a straightforward computation of its norm would grow quadratically in the number of summands of the decomposition.
By contrast, the procedure here, called the fast norm estimation, computes the norm of a sparsified state with a cost that scales linearly rather than quadratically in the number of Gaussian states in the superposition.
This allows for an overall linear scaling in the $\chi$ ---the number of Gaussian states in the decomposition---for the non-Gaussian simulator.
In the following, we present the procedure of fast norm estimation for non-Gaussian states. A more detailed derivation can be found in Appendix~\ref{ap:fastnorm}.

It is known that coherent states and displacement operators in CV cases can be used analogously to the 1-design of multiqubit cases.
We can represent the identity using coherent states or displacement operators as
\begin{align}
    \mathds{1}= \frac{1}{\pi} \int_\mathds{C} \dd \alpha\,  \dyad{\alpha}=\frac{1}{\pi} \int_\mathds{C} \dd\alpha\,  D(\alpha)\dyad{0}D^\dag(\alpha).
\end{align}
This means that, using a uniformly weighted integral over all coherent states $\ket{\bm{\xi}} = D(\bm{\xi})\ket{0}$
for $X$ defined as
\begin{align}
    X\coloneqq\pi^{-n} \abs{ \bra{\bm{\xi}}\ket{\Omega}}^2,
\end{align}
we have that
\begin{equation}
\int_{\mathds{C}^n}\dd\bm{\xi}\,X= \braket{\Omega},   
\end{equation}
where $D(\bm{\xi})$ is integrated over the non-weighted choices of $\bm{\xi}$.
Consequently, if we can approximate this integral by some quadrature, computing the overlap between a random coherent state and the state $\ket{\Omega}$ can be accomplished in time scaling linearly with the number of Gaussian states in the decomposition of $\ket{\Omega}$.

Yet problematically, the distribution of all displacements is not compact, and it is therefore impossible to sample from a uniform distribution over it.
To resolve this problem, similar to Ref.~\cite{zhuang2019scrambling}, we use sampling from a Gaussian ensemble
\begin{align}
\label{eq:Gaussian_ens}
    \mathcal{D}_N=\left\{ D(\bm{\xi}):\bm{\xi} \sim P_D^G(\bm{\xi},N)=\frac{e^{-\abs{\bm{\xi}}^2/N}}{\pi^n N^n} \right\},
\end{align}
which reproduces the identity in the limit of
\begin{align}
\label{eq:1-design}
    \lim_{N\rightarrow\infty}N^n\int_{\mathds{C}^n} \dd\bm{\xi}\,   P_D^G(\bm{\xi},N) D(\bm{\xi})\dyad{0} D^\dagger(\bm{\xi})=\mathds{1},
\end{align}
and it also holds that 
\begin{align}
\label{eq:limit_2}
&\lim_{N\rightarrow\infty}N^n\int_{\mathds{C}^n} \dd\bm{\xi}\,   P_D^G(\bm{\xi},N) D(\bm{\xi})^{\otimes 2}\dyad{0\otimes 0} D^\dagger(\bm{\xi})^{\otimes 2}\nonumber\\
&\quad = \frac{1}{\pi^n}\int_{\mathds{C}^n} \dd\bm{\xi} D(\bm{\xi})^{\otimes 2}\dyad{0\otimes 0} D^\dagger(\bm{\xi})^{\otimes 2}.
\end{align}
In particular, these limits mean that,
for any $\delta>0$, there is a sufficiently large $N_\delta$ such that, for a state $\ket{\Omega}$ of interest,~\footnote{
A rigorous mathematical proof of the interchange of limit and integral in Eqs.~\eqref{eq:1-design} and~\eqref{eq:limit_2} is left for future work. The scope of this paper is a physical regime where Eqs.~\eqref{eq:1-design} and~\eqref{eq:limit_2} can be assumed, as in Ref.~\cite{zhuang2019scrambling}.
} 
\begin{equation}
\label{eq:Gaussian expected value}
\begin{aligned}
  &\left|N_\delta^n \int_{\mathds{C}^n} \dd \bm{\xi} P_D^G(\bm{\xi},N_\delta) \bra{\Omega}D(\bm{\xi}) \dyad{0} D^\dagger(\bm{\xi})\ket{\Omega}-\braket{\Omega}\right|\\
  &\quad\leq \delta\braket{\Omega},
\end{aligned}\end{equation}
and
\begin{widetext}
\begin{align}
\label{eq:Gaussian variance}
    &\Big|\pi^n N_\delta^{n}  \int_{\mathds{C}^n} \dd \bm{\xi} P_D^G(\bm{\xi},N_\delta) \bra{\Omega \otimes \Omega} D(\bm{\xi})\otimes D(\bm{\xi}) \dyad{0\otimes 0} D^\dagger(\bm{\xi})\otimes D^\dagger(\bm{\xi})\ket{\Omega \otimes \Omega}\nonumber\\
    &\quad-\int_{\mathds{C}^n} \dd\bm{\xi} \bra{\Omega \otimes \Omega}(D(\bm{\xi})\otimes D(\bm{\xi}))\dyad{0\otimes 0} (D^\dagger(\bm{\xi})\otimes D^\dagger(\bm{\xi}))\ket{\Omega \otimes \Omega}\Big|\leq \delta\braket{\Omega \otimes \Omega}.
\end{align}
\end{widetext}

Based on Eq.~\eqref{eq:Gaussian expected value}, we define the random variable as 
\begin{align}
    X=N_\delta^n \abs{\bra{\bm{\xi}}\ket{\Omega}}^2.
\end{align}
Thus, if we sample random displacements or, equivalently, coherent states from the ensemble in Eq.~\eqref{eq:Gaussian_ens} with $N=N_\delta$ and compute the overlap with the state of interest $\ket{\Omega}$,
we can estimate the norm of $\ket{\Omega}$ within the desired target error $\delta$.

In order to bound the number of samples needed to estimate the norm, we will use Chebyshev's inequality and thus need to bound the variance of $X$.
Due to~\eqref{eq:Gaussian variance}, the variance is upper-bounded by
\begin{widetext}
\begin{align}
    \text{Var}[X]&\leq \mathds{E}[X^2]\\
    &= N_\delta^{2n}  \int_{\mathds{C}^n} \dd \bm{\xi} \frac{e^{-\abs{\bm{\xi}}^2/N_\delta}}{\pi^n N_\delta^n} \bra{\Omega \otimes \Omega} D(\bm{\xi})\otimes D(\bm{\xi}) \dyad{0\otimes 0} D^\dagger(\bm{\xi})\otimes D^\dagger(\bm{\xi})\ket{\Omega \otimes \Omega}\\
    &\leq \frac{N_\delta^{n} } {\pi^{n}} 2^{-n} \bra{\Omega \otimes \Omega} \Pi \ket{\Omega \otimes \Omega}+ \delta \braket{\Omega\otimes \Omega}\\
    &\leq \frac{2^{-n}N_\delta^{n}+\delta\pi^n}{\pi^{n}} \norm{\Omega}^4.
\end{align}
where $\Pi$ is a projector.
\end{widetext}

Thus, by performing IID sampling of $L$ coherent states from the Gaussian ensemble in Eq.~\eqref{eq:Gaussian_ens}, we can estimate the overlap with the state $\ket{\Omega}$ for sufficiently large $N_\delta$.
In particular, we can define the estimator as
\begin{align}
    \eta = \frac{1}{L}\sum_{i=1}^L N_\delta^n \abs{ \bra{\alpha_i}\ket{\Omega}}^2,
\end{align}
where $\ket{\alpha_i}$ is the coherent state for the $i$th sampling.
Then, $\eta$ estimates the expectation value $\norm{\Omega}^2$ and has a variance that is upper bounded by $\sigma^2 \leq L^{-1}(2^{-n}N_\delta^{n}+\delta/\pi^n)/\pi^n \norm{\Omega}^4$.
Using Chebyshev's inequality and choosing the number of samples as
\begin{equation}
    L=\frac{2^{-n} N_\delta^{n}+\delta \pi^n}{\pi^n}  \epsilon^{-2} p_f^{-1},
\end{equation}
we can conclude that the estimator is bounded, with a probability of at least $1-p_f$, by
\begin{align}
    (1-\epsilon-\delta)\norm{\ket{\Omega}}^2 \leq \eta \leq (1+\epsilon+\delta)\norm{\ket{\Omega}}^2.
\end{align}
See also Appendix~\ref{ap:fastnorm} for more details.

The bounds we investigated can be improved using physical insights in the states we want to simulate.
By taking into account the mean photon number of the sparsified state $\ket{\Omega}$, that is $ N_\Omega=\frac{\bra{\Omega}n\ket{\Omega}}{\braket{\Omega}} =\frac{1}{\pi^n \braket{\Omega}} \int_{\mathds{C}^n} \dd \bm{\xi} \abs{\xi}^2 \abs{\bra{\xi}\ket{\Omega}}^2$, we can improve the fast norm algorithm.
See Appendix~\ref{ap:refined} for more details.
By explicitly using the mean photon number $N_{\Omega}$ of the state $\ket{\Omega}$, we can bound the expectation value of $X$ by
\begin{align}
    \braket{\Omega}\qty(1-\frac{N_\Omega}{N})\leq\mathds{E}(X)\leq \braket{\Omega},
\end{align}
and the variance by
\begin{align}
    \text{Var}[X]&\leq \mathds{E}[X^2]\leq\frac{N^{n} } { 2^n \pi^{n}}  \norm{\Omega}^4.
\end{align}
Using the same techniques as before, we obtain that with using 
\begin{align}
  L=\frac{2^{-n} N_\delta^{n}}{\pi^n}  \epsilon^{-2} p_f^{-1}
\end{align}
samples the estimator $\eta$ is bounded with a probability of at least $1-p_f$ by 
\begin{align}
    (1-\epsilon-\frac{N_{\Omega}}{N})\norm{\ket{\Omega}}^2 \leq \eta \leq (1+\epsilon)\norm{\ket{\Omega}}^2.
\end{align}
In the end, the fastnorm algorithm scales linearly in the number of terms required to represent the non-Gaussian state; however, the runtime now depends on the mean photon number of the state of interest.

Note that whereas we presented our non-Gaussian simulator for pure states, the simulation works for the mixed states with minor modifications; that is for a state $\rho=\sum_i p_i \dyad{\psi_j}$ one starts by sampling $\ket{\psi_j}$ with  $p_i$ and then runs the simulation for the pure state $\ket{\psi_j}$.

\subsection{Comparison}

In this subsection, we compare the algorithms presented in this manuscript with other state-of-the-art methods.

The most commonly used simulation technique to treat infinite-dimensional systems is based on Fock space expansions. These methods are very versatile since states and operators can be expressed as matrices, and then standard linear algebra tools can be applied. The big downside with this approach is that the matrices tend to get large, especially when one needs to treat multi-mode systems. Using this approach, even Gaussian states become intractable very quickly. Additionally, one makes approximations from the start of the simulation. 
Our simulation algorithm circumvents these problems. The algorithm in this manuscript can always treat Gaussian states efficiently, and there is no \textit{a priori} increase in simulation complexity by considering more modes.

The algorithm presented in~\cite{Marshall2023simulation} is similar to the exact algorithm presented in this manuscript.
Reference~\cite{Marshall2023simulation} considers decompositions in coherent states and allows passive Gaussian unitaries. Our algorithms allow for a larger class of operations that do not increase the simulation cost. Furthermore, the decomposition of states using Gaussian states instead of coherent states will lead to the same or smaller simulation cost since coherent states are a subset of Gaussian states. Furthermore, the approximate simulation algorithm leads to an improved scaling to linear in the number of terms in the decomposition.

\section{Measures of non-Gaussianity based on Gaussian decomposition}
\label{sec:measures}

It is known that the coherent states and, therefore, the Gaussian state form an over-complete basis. The references~\cite{BARGMANN1971221, Perelomov1971-rb} study which subset of the coherent states form a complete basis.
With such a basis, we can expand an arbitrary state $\ket{\psi}$ as 
\begin{align}
    \ket{\psi}=\sum_{m,n} c_{m,n}\ket{\alpha_{m,n}}
\end{align}
 for a coherent state $\ket{\alpha_{m,n}}$ with $\alpha= m \omega_1 +n \omega_2$, where the complex numbers $\omega_1, \omega_2$ span a cell with area $\pi$. The simplest example is $\alpha_{m,n}=\sqrt{\pi}\qty(m+i n)$. This countably infinite subset of the coherent state is complete. 
Thus, one can, in principle, express a quantum state $\ket{\psi}$ in a countable infinite basis. 
We relax this notation and not only optimize over the set of coherent states, but arbitrary Gaussian states. 
Since the set of coherent states is a proper subset of Gaussian states, expanding states in a superposition of Gaussian states is possible, but may be impractical for some states/hard to find.

We can therefore represent any arbitrary state as a sum of Gaussian states
\begin{align}
    \ket{\psi} = \sum_{k=1}^\chi c_k\ket{\sigma_k,\bm{\mu_k}}
\end{align}
where we label each Gaussian state with covariance $\sigma_k$ and mean $\bm{\mu_k}$, and $\chi\in\{1,2,\ldots,\infty\}$.
We have seen before that the exact simulation algorithms scale with $\chi^2$, while the approximate algorithm linearly in $\norm{c}_1^2$. Decomposing arbitrary states in superposition of Gaussian states allows us to define quantifiers of non-Gaussianity as in the framework of quantum resource theories~\cite{RevModPhys.91.025001,Kuroiwa2020generalquantum}.
In analogy to the resource theory of magic~\cite{bravyi2016improved,PhysRevX.6.021043,bravyi2019simulation}, we define the Gaussian rank as
\begin{align}
\label{eq:rank}
    \chi(\ket{\psi})\coloneqq\inf\qty{\chi: \ket{\psi} = \sum_{k=1}^\chi c_k\ket{\sigma_k,\bm{\mu_k}}},
\end{align}
and the Gaussian extent as
\begin{align}
\label{eq:extent}
    \xi(\ket{\psi})\coloneqq \inf\qty{  \norm{c}_1^2:\ket{\psi} = \sum_{k} c_k\ket{\sigma_k,\bm{\mu_k}}  }.
\end{align}
We allow in the definition of the Gaussian rank and the Gaussian extent countably infinitely many terms in the decomposition.
A decomposition of quantum states in coherent states was instead considered in the context of the resource theory of non-classicality~\cite{PhysRevA.90.033812}.

Sometimes it is more convenient to decompose the state in continuous decomposition instead of sums. We have seen that the approximate simulation algorithm works equivalently if one chooses a continuous decomposition. Thus we can
use the integral in place of the sum in the definition of these measures to take into account continuous decomposition, i.e.,
\begin{align}
    \xi^\prime(\ket{\psi})\coloneqq \inf\qty{  \norm{c}_1^2:\ket{\psi} = \int{dk}\, c_k\ket{\sigma_k,\bm{\mu_k}}  }.
\end{align}
We leave the analysis of such measures for future work.
The Gaussian extent can always be used to upper-bound the approximate Gaussian rank, which is defined as 
\begin{align}    \chi_\delta(\ket{\psi})\coloneqq\inf\qty{\chi(\ket{\psi'}): \norm{\psi -\psi'}<\delta}.
\end{align}
The upper bound is then given as 
\begin{align}
     \chi_\delta(\ket{\psi})\leq 1+\frac{ \xi(\ket{\psi})}{\delta^2},
\end{align}
which follows from Eq.~\eqref{eq:sparcification} that extends the argument in Ref.~\cite{bravyi2019simulation} for magic states to non-Gaussian states.
The two measures for the resource theory of non-Gaussianity we have introduced here have a powerful operational meaning; namely, they quantify the simulation overhead introduced by the non-Gaussian nature.  
These measures can be naturally extended to density matrices, e.g.,
\begin{align}
\label{eq:mixed_state_measure}
    \Xi(\rho)=\inf \qty{ \sum_j p(j) \xi(\ket{\psi_j}): \rho=\sum_j p(j) \dyad{\psi_j}  },
\end{align}
where the minimization is taken over all possible ensembles $\{p(j),\ket{\psi_j}\}$ such that $\rho= \sum_j p_j \dyad{\psi_j}$ and $\dyad{\psi_j}$ are pure quantum states as in the conventional technique called convex roof extentions~\cite{RevModPhys.91.025001}.

It is possible to use the continuous decomposition as well, i.e.,
\begin{align}
    \Xi^\prime(\rho)=\inf \qty{ \int{dj}\,  p(j) \xi(\psi_j): \rho=\int{dj}\, p(j) \dyad{\psi_j}  },
\end{align}
while we leave the analysis of the difference in these definitions for future work.
It is also straightforward to define the approximate version of these measures in the same way as those in the resource theory of magic~\cite{bravyi2016improved,bravyi2019simulation}.

These functions are valid measures of non-Gaussianity satisfying the monotonicity under Gaussian operations as follows.
To see the monotonicity, it suffices to check that the composition with Gaussian states, applications of Gaussian unitaries, and measurement by heterodyne or homodyne detection cannot increase the Gaussian rank and the Gaussian extent. 
It is easy to see that the composition with Gaussian states as well as application of Gaussian unitaries do not increase the Gaussian rank or extent, because of the fact that the Gaussian states in the decomposition will be mapped to a different Gaussian state, while the number of Gaussian states in the superposition is left invariant.
What is left to check is monotonicity under heterodyne measurements.
Heterodyne measurements on subsystem B consisting of $m$ modes are given by the following POVMs $\Pi_{\dyad{\bm{\alpha}}}=\frac{1}{(2\pi)^m}\dyad{\bm{\alpha}}$.
In order to improve the readability, we will use the notation $\ket{G_k}=\ket{\sigma_k,\bm{\mu_k}}$ for the Gaussian state $k$.
By choosing one specific outcome, we get
\begin{align}
    \ket{\psi_{\bm{\alpha}}}&=\frac{1}{\sqrt{p(\bm{\alpha})}} \Pi_{\dyad{\bm{\alpha}}}^B\ket{\psi}\\
    &= \sum_{k=1}^\chi \frac{c_k}{\sqrt{p(\bm{\alpha})}} \Pi_{\dyad{\bm{\alpha}}}\ket{G_k}^{AB} \\
    &=\sum_{k=1}^\chi \frac{c_k^{\bm{\alpha}}}{\sqrt{p(\bm{\alpha})}} \ket{G_k^{\bm{\alpha}}}^{AB},
\end{align}
where the covariance matrix gets changed according to Sec.~\ref{sec:background} and $p(\bm{\alpha})=\Tr [\Pi_{\dyad{\bm{\alpha}}}^B\dyad{\psi}]$, $c_k^{\bm{\alpha}}=c_k \sqrt{\Tilde{p}_k(\bm{\alpha})}$, $\ket{G_k^{\bm{\alpha}}}^{AB} = \Pi_{\dyad{\bm{\alpha}}}\ket{G_k}^{AB} / \sqrt{\Tilde{p}_k(\bm{\alpha})}$ with $\Tilde{p}_k(\bm{\alpha})=\Tr [\Pi_{\dyad{\bm{\alpha}}}^B\dyad{G_k^{AB}}]$.
Therefore, we see that the Gaussian rank of the post-measurement state is upper-bounded by the Gaussian rank of the pre-measurement state. A similar argument can be given homodyne and other Gaussian measurements.

Furthermore, we have for the Gaussian extent that
\begin{align}
     \xi(\ket{\psi})\geq \int_{\mathds{C}^m } \dd \bm{\alpha} p(\bm{\alpha}) \xi(\ket{\psi_{\bm{\alpha}}}).
\end{align}
We can see this as follows.
\begin{widetext}

The Gaussian extent $\xi$ in \eqref{eq:extent} can be written by 
\begin{equation}\begin{aligned}
\xi(\ket{\psi})
&= \inf \lset \left(\sum_i |c_i|\right)^2  \sbar \ket{\psi} = \sum_i c_i \ket{\phi_i},\ c_i\in\mathbb{C},\ \ket{\phi_i}\in f \rset\\
&= \inf \lset \left(\sum_i c_i\right)^2  \sbar \ket{\psi} = \sum_i c_i \ket{\phi_i},\ c_i\geq 0,\ \ket{\phi_i}\in f\rset \\
&= \inf \lset \mu^2  \sbar \ket{\psi} = \mu \sum_i p_i \ket{\phi_i},\ \ket{\phi_i}\in f \rset\\
&= \inf \lset \mu^2  \sbar \ket{\psi}\in\mu\ \conv f \rset 
\label{eq:extent base norm}
\end{aligned}\end{equation}
where in the second line we used the fact that $e^{i\theta}\ket{\phi}\in \mathcal{G}$ if $\ket{\phi}\in\mathcal{G}$, and in the third line $\{p_i\}_i$ denotes a probability distribution.
\end{widetext}
This means that the optimal decomposition of a state $\ket{\psi}$ is given as 
\begin{align}
    \ket{\psi} = \mu \sum_i p_i \ket{G_i}.
\end{align}
We will write $\ket{\sigma} = \sum_i p_i \ket{G_i} \in  \conv \mathcal{G}$, meaning that $\ket{\sigma}$ is in the convex superposition of Gaussian states with $\sum_i p_i=1$.
By applying a heterodyne measurement $\Pi_{\dyad{\bm{\alpha}}}$ or, more generally, a free completely positive map, we know that
\begin{align}
    \Pi_{\dyad{\bm{\alpha}}}\ket{\sigma}&= \sum_i p_i \Pi_{\dyad{\bm{\alpha}}} \ket{G_i}\\
    &=\sum_i p_i  \ket{G_i^{\bm{\alpha}}} =:\eta_{\bm{\alpha}},
\end{align}
where $\ket{G_i^{\bm{\alpha}}}= \Pi_{\dyad{\bm{\alpha}}} \ket{G_i}$. 
Then, $\eta_{\bm{\alpha}}$ is in $\eta_{\bm{\alpha}} \in \norm{\eta_{\bm{\alpha}}} \conv \mathcal{G} =\norm{\Pi_{\dyad{\bm{\alpha}}}\ket{\sigma}} \conv \mathcal{G}.$
So we can write that
\begin{align}
    \Pi_{\dyad{\bm{\alpha}}} \ket{\psi}= \mu \eta_{\bm{\alpha}} = \mu \norm{\eta_{\bm{\alpha}}} \ket{\bar{G}}
\end{align}
where $\ket{\bar{G}}\in \conv \mathcal{G}$.
This is a feasible solution to $\xi(\frac{\Pi_{\dyad{\bm{\alpha}}} \ket{\psi}}{\sqrt{p(\bm{\alpha})}})$, however not guaranteed to be optimal and thus
\begin{align}
    \xi\qty(\frac{\Pi_{\dyad{\bm{\alpha}}} \ket{\psi}}{\sqrt{p(\bm{\alpha})}})\leq \frac{\mu^2 \norm{\eta_{\bm{\alpha}}}^2}{p(\bm{\alpha})}.
\end{align}
Furthermore we have that, since $\norm{\psi}^2=1$ and $\mu^2\geq1$ that $\norm{\sigma}\leq1$. This means however that $\int_{\mathds{C}^m } \dd \bm{\alpha} \norm{\eta_{\bm{\alpha}}} \leq 1$, since it holds that  
$ \int_{\mathds{C}^m } \dd \bm{\alpha} \Pi_{\dyad{\bm{\alpha}}} =1$.
To conclude, we get that 
\begin{align}
     \int_{\mathds{C}^m } \dd \bm{\alpha} p(\bm{\alpha}) \xi(\ket{\psi_{\bm{\alpha}}})\leq\mu^2 \norm{\eta_{\bm{\alpha}}}^2 \leq \mu^2 =  \xi(\ket{\psi}).
\end{align}

This monotonicity also holds for their extensions to mixed states due to a conventional argument for the monotonicity of convex-roof resource measures~\cite{RevModPhys.91.025001}.
It is easy to see that both measures are faithful, i.e., have the minimal value $1$ only for probabilistic mixtures of Gaussian states.

\subsection{Characterization of Gaussian extent by robustness measure}
\label{sec:robustness}
In this section, we investigate the properties of the Gaussian extent $\xi$ in Eq.~\eqref{eq:extent}.
We show a characterization of the Gaussian extent using the robustness measure. This characterization provides a powerful condition for finding optimal decomposition for the Gaussian extent.
Lower semicontinuous robustness for a set $\mF$ of states is given by~\cite[Corollary 6]{Lami2021framework}
\begin{widetext}
\bal
\underline{R}_\mF(\rho)= \sup\lset\Tr(W\rho)\sbar W\geq 0,\ \Tr(W\sigma)\leq 1,\ \forall \sigma\in\mF\rset.
\label{eq:robustness definition}
\eal
\end{widetext}
By choosing $\mathcal{F}$ as the closure of the convex hull of the set of pure Gaussian states~\cite{takagi2018convex,Lami2021framework}, i.e., 
$\mF = \cl\conv\{\dyad{\phi}\,|\,\ket{\phi}\in\mathcal{G}\}$, \eqref{eq:robustness definition} represents the lower semicontinuous robustness of non-Gaussianity. 

To see the relation to the Gassian extent, let $\bar{\xi}$ be the quantity defined by
\bal
 \bar{\xi} (\ket{\psi})\coloneqq \inf \lset \mu^2  \sbar \ket{\psi}\in\mu\ {\rm cl}\conv \mathcal{G} \rset
 \label{eq:extent base norm closure}
\eal
 where $\mu\ {\rm cl}\conv \mathcal{G}={\rm cl}\conv \{\mu\ket{\phi}\mid \ket{\phi}\in\mathcal{G}\}$ is the closure of the convex hull of vectors in $\mathcal{G}$ scaled by $\mu$.

Comparing \eqref{eq:extent base norm closure} and \eqref{eq:extent base norm} implies $\bar{\xi}(\ket{\psi})\leq \xi(\ket{\psi})$. Together with the identification between $\bar{\xi}(\ket{\psi})$ and $\underline{R}_\mF(\dyad{\psi})$~\cite[Proposition 17]{Lami2021framework}, we get 
\begin{equation}
    \xi(\ket{\psi})\geq \bar{\xi}(\ket{\psi})=\underline{R}_\mF(\dyad{\psi})
\end{equation}
for an arbitrary pure state $\psi$.

This means that if we find a decomposition $\ket{\psi}=\sum_i c_i \ket{\phi_i}$ with $\ket{\phi_i}\in\mathcal{G}$ such that $(\sum_i|c_i|)^2$ matches $\underline{R}_\mF(\dyad{\psi})$, we can conclude that $ \xi(\ket{\psi})=\bar{\xi}(\ket{\psi})=\underline{R}_\mF(\dyad{\psi})$ for such a state. 

On the other hand, we show in Appendix~\ref{ap:witness decomposition} that whenever the equality $\xi(\psi)=\underline{R}_\mF(\dyad{\psi})$ holds and the optimal values for both expressions are achieved with a certain decomposition and a witness operator, the optimal decomposition $\ket{\psi}=\sum_i c_i\ket{\phi_i},\,\ket{\phi_i}\in\mathcal{G}$ such that $(\sum_i |c_i|)^2=\xi(\ket{\psi})$ is related to the optimal witness operator $W$ for $\underline{R}(\dyad{\psi})$ satisfying $\underline{R}_\mF = \Tr(W\dyad{\psi})$ as   
\bal
 |\Tr(W \dyad{\phi_i})| = 1, \forall i.
 \label{eq:witness decomposition}
\eal
We prove this for general resource theories defined in infinite-dimensional Hilbert spaces equipped with an arbitrary subset of pure states, extending the corresponding relation established for finite-dimensional theories in the context of the resource theory of magic~\cite[Lemma 3]{Heimendahl2021stabilizerextentis}.

This characterization is a useful guide to finding the optimal decomposition of a state for which we know the optimal witness for robustness.
For instance, we know for Fock state $\ket{n}$ that~\cite[Proposition 28]{Lami2021framework}
\bal
 \underline{R}_\mF(\ketbra{n}) = \frac{1}{\sup_{\alpha,\xi} |\langle n |\alpha,\xi\rangle|^2},
 \label{eq:sc-robustness}
\eal
where $\xi$ is a squeezing parameter. 
Letting $\ket{\alpha^\ast, \xi^\ast}$ be a state that optimizes the right-hand side, we have 
\bal
 W = \frac{\ketbra{n}}{|\langle n | \alpha^\ast,\xi^\ast \rangle|^2}.
\eal

In turn, the free states in the optimal decomposition would need to satisfy 
\bal
 |\langle n | \phi_i \rangle|^2 = |\langle n | \alpha^\ast,\xi^\ast \rangle|^2,\ \forall i.
 \label{eq:optimal concdition}
\eal
These states include $\ket{e^{i\theta} \alpha^\ast,  \xi^\ast}$, i.e., the states obtained by a phase shift from the seed state $\ket{\alpha^\ast, \xi^\ast}$. 
As we will see in the next section (Sec.~\ref{sec:applications}), the optimal decomposition fulfills these requirements.

Finding optimal $\alpha^\ast$ and $\xi^\ast$ involves optimization in general.
Nevertheless, for the Fock state of $n=1$, we have an analytical form~\cite[Proposition 29]{Lami2021framework}
\bal
\label{eq:optimal_alpha_xi}
 \alpha^\ast = \frac{2}{3},\ \xi^\ast = \ln \sqrt{3}.
\eal

\section{Decompositions and Applications}
\label{sec:applications}
\subsection{Optimal Decompositions for the Gaussian extent}
In this section, we give Gaussian decompositions for states that are of interest to the CV quantum computing community and show optimality for some of those.

The decomposition for the Fock state $\ket{n=1}$ is obtained as follows.
In the previous section, we saw that the optimal value for the Gaussian extent of the Fock state $\ket{1}$ is given with the parameters in Eq.~\eqref{eq:optimal_alpha_xi}.
We can construct the corresponding Gaussian decomposition in the following way.
For this purpose, we use the projectors defined in Ref.~\cite{grimsmo2020quantum}, i.e.,
\begin{align}
    \Pi^l_{2N}&=\sum_{k=0}^\infty \dyad{2kN+l}\\
    &=\frac{1}{2N} \sum_{m=0}^{2N-1} \qty(e^{i\pi l/N} e^{i\pi n/N}  )^m,
\end{align}
where $n$ is the photon number operator.
In particular, we define a projector into the Fock state $\ket{1}$ as
\begin{align}
    \dyad{1}=\lim_{N\rightarrow\infty} \Pi^1_{2N}.
\end{align}
Then, we apply this projector to a Gaussian state 
\begin{align}
    \lim_{N\rightarrow\infty} \Pi^1_{2N} \ket{G} = \bra{1}\ket{G} \ket{1} .
\end{align}
We need a normalization to get $\ket{1}$. In order to minimize $\norm{c}_1$, we choose the Gaussian state that has the maximum overlap with the Fock state $\ket{1}$, since it in turn will minimize $\norm{c}_1$.
Among coherent states, the maximum overlap is achieved by $\ket{\alpha=1}$ with $\bra{1}\ket{\alpha=1}=\frac{1}{\sqrt{e}}$. Intuitively, the operator
$\qty(e^{i\pi l/N} e^{i\pi n/N}  )^m$ is a phase-shifting operator. In the limit $\lim_{N\rightarrow\infty} \Pi^1_{2N}$, it superposes phase-shifted versions of the same state $\ket{G}$ on a circle with the same weight. 
We can immediately read the coefficients
\begin{align}
   \ket{1}&= \frac{1}{ \bra{1}\ket{G} } \lim_{N\rightarrow\infty} \Pi^1_{2N} \ket{G}\\
   &= \frac{1}{ \bra{1}\ket{G} }   \lim_{N\rightarrow\infty}  \frac{1}{2N} \sum_{m=0}^{2N-1} \qty(e^{i\pi l/N} e^{i\pi n/N}  )^m \ket{G},  
\end{align}
then we have $\norm{c}_1 = \sqrt{e}$. We see that all coherent states in the decomposition have the same weight.
For the more general Gaussian states, the state that maximizes the overlap is  $\ket{\alpha=\frac{2}{3},\xi=\ln \sqrt{3}}$. The overlap between the state $\ket{\alpha=\frac{2}{3},\xi=\ln \sqrt{3}}$ and the Fock state $\ket{1}$ is $\abs{\bra{1} \ket{\alpha=\frac{2}{3},\xi=\ln \sqrt{3}}}^2= \frac{3\sqrt{3}}{4 e}$. 
Using the state $\ket{\alpha=\frac{2}{3},\xi=\ln \sqrt{3}}$ as the seed state $\ket{G}$, we obtain the decomposition of Fock state $\ket{1}$ with $\norm{c}_1^2=\frac{4e}{3\sqrt{3}}$.
By comparing with the reported optimal result in~\cite{Lami2021framework}, we see that this decomposition is optimal.
To summarize, the optimal decomposition of $\ket{1}$ is then a superposition of the seed state $\ket{\alpha=\frac{2}{3},\xi=\ln \sqrt{3}}$ on which phase-shift operators have been applied to. This decomposition fulfills Eq.~\eqref{eq:optimal concdition}, e.g. it holds that for all $m$ that
\begin{align}
    &\abs{\bra{1}\qty(e^{i\pi l/N} e^{i\pi n/N}  )^m  \ket{\alpha=\frac{2}{3},\xi=\ln \sqrt{3}}}^2\\
    &= \abs{\bra{1} \ket{\alpha=\frac{2}{3},\xi=\ln \sqrt{3}}}^2. 
\end{align}

A family of states that have many applications for CV quantum computing~\cite{konno2024logical, hahn2022quantifying,PRXQuantum.6.010330}, especially for error correction, is GKP states~\cite{PhysRevA.64.012310}. A finite-energy GKP state that encodes a computational basis state $\mu$ of a $d$-dimensional qudit can be decomposed into a sum of squeezed states by definition, i.e.,~\cite{PhysRevA.64.012310,PhysRevA.102.032408}
\begin{align}
    \ket{\mu_{\kappa, \Delta}} &=\frac{1}{\sqrt{\mathcal{N}}}\sum_{s=-\infty}^\infty e^{-\frac{1}{2}\kappa^2 \alpha_d^2(ds+\mu)^2}\\
    &\times D\qty(\alpha_d(ds+\mu))S(-\log \Delta) \ket{0}\\
    &=\frac{1}{\sqrt{\mathcal{N}}}\sum_{s=-\infty}^\infty e^{-\frac{1}{2}\kappa^2 \alpha_d^2(ds+\mu)^2}\ket{G_s},
\end{align}
where $\mathcal{N}$ is a normalization constant. The constant $\kappa^{-1}$ is the width of the Gaussian envelope, $\Delta$ describes the  individual squeezing parameter with $\log(\Delta^{-1})$ and $\alpha_d=\sqrt{\frac{2\pi }{d}}$ a constant. One recovers the ideal GKP state for $\Delta,\kappa \rightarrow0$.
We use the notation $G_s= D\qty(\alpha_d(ds+\mu))S(-\log \Delta) \ket{0}$  to show that the GKP state can be written as a superposition of Gaussian states.

We see that states labeled by high $s$ are exponentially suppressed in the superposition, making the simulator based on the Gaussian extent significantly more viable as compared to the one based on the Gaussian rank.

Another family of bosonic codes is the cat code. The cat code is an example of a rotational symmetric code~\cite{grimsmo2021quantum}.
The codewords of the cat code are written as
\begin{align}
    \ket{\mu=0/1^M}=\frac{1}{\sqrt{\mathcal{N}}}\sum_{m=0}^{2M-1} (-1)^{\mu m} e^{i \frac{m\pi}{M}n}\ket{\alpha},
\end{align}
where $\ket{\alpha}$ is a coherent state, and $\mathcal{N}$ is a normalization constant.
For squeezed cat codes, the superposition goes over squeezed states instead of coherent states.
The most commonly used states of this family are the even/odd cat states
\begin{align}
\label{eq:cat}
    \ket{\alpha_\pm}=\frac{1}{\sqrt{\mathcal{N}}}\qty(\ket{\alpha}\pm \ket{-\alpha})
\end{align}
with $\mathcal{N}=2(1\pm e^{-2\abs{\alpha}^2})$.
For this decomposition, we obtain that $\norm{c}_1= \frac{2}{\sqrt{\mathcal{N}}}=\sqrt{\frac{2}{1\pm e^{-2\abs{\alpha}^2}}}$.
It trivially holds that these decompositions are optimal in minimizing the Gaussian rank in Eq.~\eqref{eq:rank} for non-vanishing $\alpha$  because $2$ is the smallest rank for a non-Gaussian state.

The optimal decomposition that minimizes the Gaussian extent in Eq.~\eqref{eq:extent} requires more careful analysis.
For the resource theory of non-classicality, the decomposition Eq.~\eqref{eq:cat} is optimal for the case of $\ket{\alpha_+}$; for $\ket{\alpha_-}$, the same is the case for approximately $\alpha>1$.
Both cases yield $\norm{c}_1=\sqrt{2}$ for $\alpha\rightarrow \infty$.
For the resource theory of non-Gaussianity, the previous decomposition is an upper bound, and the optimality is unknown in general.
To gain further insight, consider the extreme cases $\lim_{\alpha \rightarrow 0 }$ $\ket{\alpha_+}=\ket{0}$ and $\lim_{\alpha \rightarrow 0 } \ket{\alpha_-}=\ket{1}$, which yield $\norm{c}_1^2=1$ and $\norm{c}_1^2= \frac{4e}{3\sqrt{3}}$, respectively.
In the opposite case of $\alpha \rightarrow \infty$, it should converge to $\norm{c}_1=\sqrt{2}$ even if one exchanges the coherent states with squeezed states (squeezed along the correct axis) as for the non-classicality case; after all, they asymptotically become orthogonal.
Therefore, the decomposition above is optimal for large enough cat states (and/or squeezing).
It could even be that for $\ket{\alpha_+}$, the optimal decomposition is the same for non-Gaussianity and non-classicality.
Note here that either way, the scaling is better than the one proposed in~\cite{bourassa2021fast}, reducing the simulation overhead.

In general, for non-classicality, the norm $\norm{c}_1$ and thus the optimal decomposition is multiplicative for single-mode states~\cite[Proposition 24]{Lami2021framework}.
However, this is not known in the case of non-Gaussianity.
In Ref.~\cite[Theorem 6.6]{dias2023classical}, it is proved that the multiplicativity of the $\mathcal{D}$-fidelity implies multiplicativity of the $\mathcal{D}$-extent for a possibly infinite dictionary $\mathcal{D}\subset \mathcal{H}$ if there exists a finite $\epsilon$-net in $\mathcal{D}$ that contains an orthonormal basis of $\mathcal{H}$. 
A sufficient condition for this is that the subset  $\mathcal{D}\subset \mathcal{H}$ is compact. However, in our case, the set of Gaussian states is \emph{not} compact, and thus their proof is not applicable to this case.

We will nonetheless report here a counterexample for the multiplicativity of the Gaussian extent. As reported before in Eq.~\eqref{eq:sc-robustness}, the lower semicontinuous robustness of non-Gaussianity is the inverse maximal fidelity of a Fock state $\ket{n}$ and pure Gaussian state $\ket{\alpha,\xi}$. Since the optimal value for $ \underline{R}_\mF(\ketbra{1})=\xi(\ket{1})= \frac{4e}{3\sqrt{3}}$, this, together with Eq.~\eqref{eq:sc-robustness}, implies that the maximal overlap between a Fock state and a Gaussian state is $\abs{\bra{1}\ket{G}}^2= \frac{3\sqrt{3}}{4e}\approx 0.47789$, where $\ket{G}$ is the Gaussian state closest to the state $\ket{1}$.
Since we do not have provable optimal results for the multimode case, we use a numerical approach. Details as well as the optimization parameters can be found in Appendix~\ref{ap:numerics}.
The highest fidelity we numerically found between the tensor product of two Fock states $\ket{1 \otimes 1}$ and a two-mode Gaussian state $\ket{G'}$ is $\abs{\bra{1\otimes 1}\ket{G^\prime}}^2=\frac{1}{4}$. 
Note here that this fidelity is not provably maximum, but is nonetheless a lower bound on the maximal fidelity between two Fock states $\ket{1 \otimes 1}$ and the set of Gaussian states.
It thus holds that 
\begin{equation}
    \abs{\bra{1\otimes 1 }\ket{G^\prime}}^2 >\abs{\bra{1}\ket{G}}^2 \abs{\bra{1}\ket{G}}^2,
\end{equation}
which shows that the Gaussian fidelity is not multiplicative for the state $\ket{1}$.
Since the Gaussian extent coincides with the Gaussian fidelity for Fock states as in Eq.~\eqref{eq:sc-robustness}, we conclude that the Gaussian extent is not multiplicative, even for single-mode states.
This means, in terms of classical simulation, that finding multi-mode expansions can improve the performance of the simulation algorithm as the simulation cost is sub-multiplicative.

\subsection{Applications}
\label{sec:bounds}
In this section, we discuss the applications of our non-Gaussian simulator and the bounds that we derive using the Gaussian extent.

\paragraph{Gaussian boson sampling:}
We start by investigating sampling models.
There are several versions of boson sampling.
All can be implemented using linear optics transformations.
Standard boson sampling~\cite{10.1145/1993636.1993682} has several Fock states as input and measures photon number
\begin{align}
    \ket{1}^N\otimes \ket{0}^{M-N} \rightarrow\text{Measure}\; n.
\end{align}
Gaussian boson sampling~\cite{PhysRevLett.119.170501} changes the input to Gaussian states and measures photon number
\begin{align}
    \ket{G}^N \rightarrow\text{Measure} \;n.
\end{align}
Equivalently, this can be reversed, and instead, we can have Fock states in the input and measure heterodyne or, more generally, a Gaussian measurement~\cite{Chabaud2017continuous, PhysRevResearch.3.033018}
\begin{align}
    \ket{1}^N\rightarrow\text{Measure}\; G.
\end{align}
All variants of boson sampling are based on the fact that it is hard to compute the output probabilities of measuring a certain photon pattern in the high mode regime, where the number of modes scales faster than the number of photons.

Among these variants, Gaussian boson sampling and the reversed variant are especially suited for our approach.
The probability of measuring a certain output pattern in a Gaussian boson sample experiment is
\begin{align}
    P(\Vec{n})=\Tr\qty[\bigotimes_{j=1}^M \dyad{n_j}U\dyad{G}U^\dagger],
\end{align}
where we start with a Gaussian pure state $\dyad{G}$ and then evolve it under a Gaussian (passive) unitary $U$ with $n_j=0/1$.
This is equivalent to the time-reversed version
\begin{align}
        P(\Vec{n}) &=\Tr\qty[\bigotimes_{j=1}^M \dyad{n_j}U\dyad{G}U^\dagger]\\
        &=\Tr\qty[U^\dagger\bigotimes_{j=1}^M \dyad{n_j}U\dyad{G}],
\end{align}
where we start with some Fock states $\ket{1}$ and $\ket{0}$, evolve them with the inverse unitary, which is again a Gaussian unitary, and then finally measure by a certain Gaussian measurement.
Interestingly, the hardness proofs require that there are more modes than photons in the system. Our simulator has, in some sense, the converse behavior. Adding Gaussian auxiliary modes in the vacuum states does not really increase the simulation cost significantly, but adding single-photon states does.

As we have shown before, the Gaussian extent is not multiplicative; therefore, we can give only an upper bound on the simulation cost and not the lowest one possible. Note that the simulation algorithm still works---it just does not have optimal performance. 
The upper bound of the simulation cost is then just $\qty(\norm{c}_1^2)^{\Bar{M}}=\qty(\frac{4e}{3\sqrt{3}})^{\Bar{M}}$, where $\Bar{M}$ is the number of Fock state $\ket{1}$ in the input. 
Therefore, our results provide an algorithm whose simulator runtime scales exponentially in the number of Fock states, allowing one to classically simulate logarithmically many Fock states efficiently. This is consistent with~\cite{PhysRevResearch.3.033018}.

It is insightful to compare the case of the resource theory of non-classicality since all the boson sampling models only involve passive symplectic unitary operations.
In this case, the corresponding extent is multiplicative, and the optimal decomposition is known with 
\bal
 \underline{R}_\mF(\ketbra{n}) = e^n \frac{n!}{n^n}.
\eal
 Therefore, to compute the probability of obtaining a specific sampling pattern that involves $\Bar{M}$ single photons, the simulation cost scales with $e^{\Bar{M}}>\qty(\frac{4e}{3\sqrt{3}})^{\Bar{M}}$. We see that even though the dynamics and measurements are fully included in the set of free operations, using the decomposition obtained for the resource theory of non-Gaussianity improves the simulation capacity, even if we only compare it with the upper bound.

\paragraph{Cat state breeding:}

The generation of grid states is of wide interest in the continuous-variable quantum information community due to their usefulness for fault-tolerant quantum computing~\cite{PhysRevA.64.012310}. The generation of such states remains challenging. One of the most promising ways is to use cat states to breed grid states~\cite{weigand2018generating}.
The breeding protocol takes in squeezed cat states and uses Gaussian operations to generate a grid state of higher quality.
The targeted grid states are the sensor states, i.e.,
\begin{align}
    \ket{\Psi_\Delta}&\propto \sum_{t=-\infty}^\infty e^{-\pi\Delta^2 t^2} D\left(t\sqrt{\frac{\pi}{2}}\right) S(\Delta)\ket{0}\\
    &\propto \sum_{t=-\infty}^\infty e^{-\pi\Delta^2 t^2} \ket{G_t}.
    \label{eq:gridstate}
\end{align}
We can use the Gaussian extent to derive a lower bound on how many cat states are needed to obtain one sensor state with a certain squeezing $\Delta$.
Cat state breeding protocols require large enough displacement in the input~\cite{weigand2018generating}, so we assume that the cat states are in the limit of near-orthogonality of $\ket{\alpha}$ and $\ket{-\alpha}$  leading to $\xi(\ket{\alpha_+})\approx 2$. We do not require this fact in order to derive a lower bound for the required copies of cat states. Since we have optimal values of the robustness for non-classicality, we have upper bounds on the Gaussian extent, which allows us to compute a lower bound on the number of copies required.
The minimal required number $n$ of cat states $\ket{\alpha_+}$ for generating a grid states $ \ket{\Psi_\Delta}$ is bounded by using the monotonicity of the Gaussian extent $\xi$ as
\begin{align}
     \xi\qty(\ket{\alpha_+^{\otimes n}})\geq \xi\qty(\ket{\Psi_\Delta}).
\end{align}
Since it holds that
\begin{align}
    \xi\qty(\ket{\alpha_+^{\otimes n}})\leq   \xi\qty(\ket{\alpha_+})^n,
\end{align}
we can find the lower bound by
\begin{align}
    n \geq \left \lceil \frac{ \log \xi(\ket{\Psi_\Delta})}{\log\xi(\ket{\alpha_+})} \right \rceil\geq  \left \lceil \frac{\log \xi(\ket{\Psi_\Delta})}{\log 2} \right \rceil.
\end{align}
For high enough squeezing, i.e., small enough $\Delta$, the Gaussian states $\ket{G_t}$ of the grid states are nearly orthogonal.
So we compute $\xi(\ket{\Psi_\Delta})$ using the coefficients of Eq.~\eqref{eq:gridstate}. The results can be found in Table~\ref{tab:cat_breeding}. These bounds are far away from the results reported in Ref.~\cite{weigand2018generating}. To obtain a squeezing level of $\Delta = 0.1$, the authors of~\cite{weigand2018generating} report $M=6$ rounds of breeding using $2^M= 64$ cat states. However, they fix the input squeezing and displacement of their cat states, while we essentially leave it indeterminate. So our bounds are valid for all input cat states, independent of their magnitude.

\begin{table}
    \centering
    \begin{tabular}{c|c c}
         \hline
         $\Delta$ &$\xi(\ket{\Psi_\Delta})$& $n$  \\\hline
           0.3 & 2.797 & 2\\
           0.2 &3.969  &2 \\
           0.1 & 7.496 & 3\\
           0.05 &14.562  & 4\\
           0.025 & 28.701 & 5\\
           0.01 & 71.126 & 7\\
         \hline
    \end{tabular}
    \caption{Numerical values of the Gaussian extent $\xi(\ket{\Psi_\Delta})$ of the grid state $\ket{\Psi_\Delta}$ with squeezing $\Delta$ in Eq.\eqref{eq:gridstate} and the lower bound on the required number $n$ of cat states for obtaining $\ket{\Psi_\Delta}$.}
    \label{tab:cat_breeding}
\end{table}

We here compare these bounds with other state-of-the-art techniques.
The most commonly studied measures in the resource theory of non-Gaussianity are the stellar rank~\cite{PhysRevLett.124.063605} and the Wigner negativity~\cite{takagi2018convex,albarelli2018resource}.
In the following, we will compare state transformation bounds that one can obtain using the stellar rank and Wigner negativity with the bounds we obtain using the Gaussian extent.
If we consider the state conversion from many cat states to one GKP state, the stellar rank cannot be used as cat and GKP states have infinite stellar rank. The Wigner negativity of cat states with large spacing is upper-bounded~\cite{Kenfack2004negativity}. The number of such cat states with large magnitude to GKP states with $\Delta=0.3,0.2,0.1$ is equivalent to the one obtained with the Gaussian extent presented in Tab.~\ref{tab:cat_breeding}.
However, if we consider the state transformation from cat states with large magnitude to a single Fock state, Wigner negativity allows this conversion. Even more, as long as the magnitude $\alpha$ is approximately larger than $\sqrt{2}$, the transformation is possible. The stellar rank allows the state conversion even for $\alpha=\epsilon$. The bounds using the Gaussian extent, however, tell us that we need at least two cat states to obtain the Fock state one.
Generally, one can say that since the Gaussian extent is sub-multiplicative and the Wigner negativity is multiplicative, the Gaussian extent will have tighter bounds in most multi-copy scenarios.

\section{Conclusion}
\label{sec:conclusion}
In this work, we introduced two efficient algorithms for the classical simulation of non-Gaussian optics. Inherently connected to these algorithms are two measures of non-Gaussianity, i.e., the Gaussian rank and the Gaussian extent, which quantify the computational cost of the simulation. 
The simulator, whose cost scales quadratically with the Gaussian rank, is exact, while the other one uses a sparsification of the non-Gaussian states and is thus approximate. This approximation allows for reducing the scaling of the simulation cost to linear in the Gaussian extent. The algorithms use extensions of the standard covariance matrix formalism that include phase-sensitive overlaps and a routine for fast norm estimation that allows for linear scaling.
Furthermore, we investigated the properties of the Gaussian extent. 
Employing the fact that the Gaussian extent is connected to the lower semicontinuous robustness of non-Gaussianity, we were able to find optimal decompositions for states that are of interest to the continuous variable quantum computing community. Using the direct connection between the Gaussian extent and the Gaussian fidelity that exists for Fock states, we give a counterexample showing that the Gaussian extent is, in general, not multiplicative.
We applied the simulator and the tools we developed in this work to boson sampling. We gave an upper bound to the simulation cost and showed that even when the dynamics are fully included within the free set of operations of the resource theory of non-classicality, it is beneficial to use the more general decompositions using Gaussian states.
At last, we made a resource-theoretic analysis of cat state breeding and derived bounds using the Gaussian extent. The bounds are fundamental lower bounds on how many cat states are required, independent of their magnitude.

Although we have focused the presentation on Gaussian circuits with input non-Gaussian states, our method might also be useful to simulate circuits with non-Gaussian operations by recasting the latter into the former with techniques similar to those used for Clifford + T circuits \cite{gheorghiu2022t, PhysRevLett.130.090602, mele2024learning}. Furthermore, in order to compute bounds on state conversions within the resource theory of non-Gaussianity, one requires provably optimal decompositions. These decompositions, in turn, improve the scaling of the simulation algorithm.
Therefore, it is an interesting path to find a provably optimal decomposition for more states that are relevant for CV quantum computing and to compute bounds. 
Our work connects the practical usefulness of an efficient classical simulator with the fundamental investigation of resources required for CV quantum computing.

\paragraph*{Note:} During the preparation of this manuscript, we became aware of related work by B. Dias and R. Koenig~\cite{dias2024classical}. We independently arrived at similar results on defining a simulator for non-Gaussian states using similar ideas to compute the overlap and estimate the norm.

\begin{acknowledgments}
 G.F. acknowledges support from the Swedish Research Council (Vetenskapsrådet) through the project grant DAIQUIRI, as well as from the HORIZON-EIC-2022-PATHFINDERCHALLENGES-01 programme under Grant Agreement Number 101114899 (Veriqub). G.F. and O.H. acknowledge support from the Knut and Alice Wallenberg Foundation through the Wallenberg Center for Quantum Technology (WACQT). 
 R.T. acknowledges the support of JSPS KAKENHI Grant Number JP23K19028, JP24K16975, JST, CREST Grant Number JPMJCR23I3, Japan, and MEXT KAKENHI Grant-in-Aid for Transformative
Research Areas A ``Extreme Universe” Grant Number JP24H00943.
 H.Y.\ was supported by JST PRESTO Grant Number JPMJPR201A, JPMJPR23FC, JSPS KAKENHI Grant Number JP23K19970, and MEXT Quantum Leap Flagship Program (MEXT QLEAP) JPMXS0118069605, JPMXS0120351339\@.
\end{acknowledgments}

\newpage
\appendix
\onecolumngrid

\section{Inner product formula}
\label{ap:innerproduct}

We present the details of the way of computing overlaps of two pure Gaussian states $\ket{G_i},\ket{G_j}\in \mathcal{G}$, i.e.,
\begin{align}
    \bra{G_i}\ket{G_j}.
\end{align}
For this purpose, we introduce a Gaussian reference state $G_0 = \dyad{G_0}$, which can be a random Gaussian state or the vacuum state.
As presented in the main text, we want to calculate overlaps between Gaussian states $\bra{G_1}\ket{G_2}$ using
\begin{align}
\label{eq:G_0G_1G_2appendix}
    \Tr \qty(G_0 G_1 G_2)&=\Tr\qty(\dyad{G_0}\dyad{G_1}\dyad{G_2})\\
    &=\bra{G_2}\ket{G_0}\bra{G_1}\ket{G_2}\bra{G_0}\ket{G_1}.
\end{align}
To obtain the desired inner product $\bra{G_1}\ket{G_2}$ from~\eqref{eq:G_0G_1G_2appendix}, we need to specify Gaussian states not only using the covariance matrix and the mean but also by specifying the inner product with the reference state.

To compute~\eqref{eq:G_0G_1G_2appendix}, we represent the Gaussian states $\ket{G}$ in terms of the characteristic function as
\begin{align}
    G= \frac{1}{(2\pi)^n}\int_{\mathds{R}^{2n}} \dd\bm{r}\, e^{-\frac{1}{4} \bm{r}^T\sigma \bm{r}+i\bm{\mu} \bm{r}} D(\Omega^T \bm{r}).
\end{align}
Using this representation, we have that
\begin{align}
    \Tr \qty(G_0 G_1 G_2) &=\frac{1}{(2\pi)^{3n}}\int_{\mathds{R}^{2n}} \dd\bm{r_0} \dd\bm{r_1} \dd\bm{\bm{r_2}}\, e^{-\frac{1}{4} \bm{r_0}^T\sigma_0 \bm{r_0}+i\bm{\mu_0} \bm{r_0}} e^{-\frac{1}{4} \bm{r_1}^T\sigma_1 \bm{r_1}+i\bm{\mu_1} \bm{r_1}}e^{-\frac{1}{4} \bm{r_2}^T\sigma_2 \bm{r_2}+i\bm{\mu_2} \bm{r_2}} \\
   &\times \Tr \qty[D(\Omega^T \bm{r_0})D(\Omega^T \bm{r_1})D(\Omega^T \bm{r_2})  ].
\end{align}
Then, using
\begin{align}
   D(\Omega^T \bm{r_1})D(\Omega^T \bm{r_2}) &= D(\Omega^T (\bm{r_1}+\bm{r_2})) e^{-\frac{i}{2} (\Omega \bm{r_1})^T \Omega (\Omega \bm{r_2})}\\
   &=D(\Omega^T (\bm{r_1}+\bm{r_2})) e^{\frac{i}{2} \bm{r_1}^T \Omega \bm{r_2}}
\end{align}
and 
\begin{align}
    \Tr\qty[D(\Omega^T r)D(\Omega^T s)]= (2\pi)^n \delta^{2n}(\Omega^T (r+s)),
\end{align}
we obtain
\begin{align}
\label{eq:A9}
    &\frac{1}{(2\pi)^{2n}}\int_{\mathds{R}^{2n}} \dd\bm{r_0} \dd\bm{r_1} \dd\bm{r_2}\, e^{-\frac{1}{4} \bm{r_0}^T\sigma_0 \bm{r_0}+i\bm{\mu_0} \bm{r_0}} e^{-\frac{1}{4} \bm{r_1}^T\sigma_1 \bm{r_1}+i\bm{\mu_1} \bm{r_1}}e^{-\frac{1}{4} \bm{r_2}^T\sigma_2 \bm{r_2}+i\bm{\mu_2} \bm{r_2}} 
    \delta^{2n}(\Omega^T(\bm{r_0}+\bm{r_1}+\bm{r_2})) e^{\frac{i}{2} \bm{r_1}^T \Omega \bm{r_2}}\\
\label{eq:A10}
    &=\frac{1}{(2\pi)^{2n}}\int_{\mathds{R}^{2n}} \dd\bm{r_0} \dd\bm{r_1}\, e^{-\frac{1}{4} \bm{r_0}^T\sigma_0 \bm{r_0}+i\bm{\mu_0} \bm{r_0}} e^{-\frac{1}{4} \bm{r_1}^T\sigma_1 \bm{r_1}+i\bm{\mu_1} \bm{r_1}}e^{-\frac{1}{4} (\bm{r_0}^T+\bm{r_1}^T)\sigma_2 (\bm{r_0}+\bm{r_1})-i\bm{\mu_2} (\bm{r_0}+\bm{r_1})} 
     e^{-\frac{i}{2} \bm{r_1}^T \Omega (\bm{r_0}+\bm{r_1})}\\
     &=\frac{1}{(2\pi)^{2n}}\int_{\mathds{R}^{2n}} \dd\bm{r_0} \dd\bm{r_1}\, e^{-\frac{1}{4} \bm{r_0}^T\sigma_0 \bm{r_0}+i\bm{\mu_0} \bm{r_0}} e^{-\frac{1}{4} \bm{r_1}^T\sigma_1 \bm{r_1}+i\bm{\mu_1} \bm{r_1}}\\
     &\times e^{-\frac{1}{4} (\bm{r_0}^T\sigma_2\bm{r_0}+\bm{r_1}^T\sigma_2\bm{r_0}+\bm{r_0}^T\sigma_2\bm{r_1}+\bm{r_1}^T\sigma_2\bm{r_1})-i\bm{\mu_2}\bm{r_0}-i\bm{\mu_2}\bm{r_1}} e^{-\frac{i}{2} \bm{r_1}^T \Omega (\bm{r_0}+\bm{r_1})}\\
\label{eq:A13}
     &=\frac{1}{(2\pi)^{2n}}\int_{\mathds{R}^{2n}} \dd\bm{r_0} \dd\bm{r_1}\, e^{-\frac{1}{4} \bm{r_0}^T\sigma_0 \bm{r_0}+i\bm{\mu_0} \bm{r_0}} e^{-\frac{1}{4} \bm{r_0}^T\sigma_2\bm{r_0}-i\bm{\mu_2}\bm{r_0}} e^{-\frac{1}{4} \bm{r_1}^T\sigma_1 \bm{r_1}+i\bm{\mu_1} \bm{r_1}}\\
     &\times e^{-\frac{1}{4} (\bm{r_1}^T\sigma_2\bm{r_0}+\bm{r_0}^T\sigma_2\bm{r_1}+\bm{r_1}^T\sigma_2\bm{r_1})-i\bm{\mu_2}\bm{r_1}} e^{-\frac{i}{2} \bm{r_1}^T \Omega \bm{r_0}}e^{-\frac{i}{2} \bm{r_1}^T \Omega \bm{r_1}}\\
\label{eq:integral}
     &=
\frac{1}{(2\pi)^{2n}} \int_{\mathds{R}^{2n}} \dd\bm{r_0} e^{-\frac{1}{4} \bm{r_0}^T\sigma_0 \bm{r_0}+i\bm{\mu_0} \bm{r_0}}  e^{-\frac{1}{4}\bm{r_0}^T\sigma_2 \bm{r_0}-i \bm{\mu_2} \bm{r_0}}   \int_{\mathds{R}^{2n}} \dd\bm{r_1}\, e^{-\frac{1}{4} \bm{r_1}^T\sigma_1 \bm{r_1}+i\bm{\mu_1} \bm{r_1}}\\
&\times e^{\frac{i}{2} \bm{r_0}^T\Omega \bm{r_1}} e^{-\frac{1}{4}\qty(\bm{r_1}^T\sigma_2 \bm{r_1}+\bm{r_1}^T\sigma_2 \bm{r_0}+ \bm{r_0}^T \sigma_2 \bm{r_1})-i\bm{\mu_2} \bm{r_1}},
\end{align}
where~\eqref{eq:A10} is obtained by using the Dirac distribution $\delta^{2n}(\Omega^T(\bm{r_0}+\bm{r_1}+\bm{r_2}))$ to replace $r_2= -r_0-r_1$, and the last line follows from the fact that $\bm{r_1}^T \Omega \bm{r_1}=0$ since $\Omega$ is anti-symmetric.
Using the Gaussian integral formula
\begin{align*}
    \int_{\mathds{R}^{2n}} \dd \bm{r}\, e^{-\frac{1}{4} \bm{r}^T A \bm{r}+\bm{B}^T  \bm{r}} = \sqrt{\frac{(2\pi)^{2n}}{\det A}}e^{\frac{1}{4}\bm{B}^T A^{-1}\bm{B}},
\end{align*}
we rewrite the integral over $\bm{r_1}$ into
\begin{align}
 \frac{1}{(2\pi)^{n}}\int_{\mathds{R}^{2n}}& \dd\bm{r_1}\, e^{-\frac{1}{4} \bm{r_1}^T (\sigma_1+\sigma_2) \bm{r_1}+ \qty(-\frac{1}{2} \bm{r_0}^T \sigma_2+i\qty[\bm{\mu_1}-\bm{\mu_2}+\frac{1}{2} \bm{r_0}^T \Omega]   )\bm{r_1}  }=\frac{1}{\sqrt{\det A}} e^{\frac{1}{4} B^T A^{-1}B},
\end{align}
where we use $\bm{r_1}^T\sigma_2 \bm{r_0}=\bm{r_0}^T\sigma_2 \bm{r_1}$, and we take
\begin{align}
    A&=\sigma_1+\sigma_2,\\
    B^T&=-\frac{1}{2} \bm{r_0}^T \sigma_2+i\qty[\bm{\mu_1}-\bm{\mu_2}+\frac{1}{2} \bm{r_0}^T \Omega].
\end{align}
The right-hand side is further simplified by
\begin{align*}
    e^{\frac{1}{4} B^T A^{-1}B}&= e^{\frac{1}{4}\qty[ (-\frac{1}{2}\bm{r_0}^T\,\sigma _2+i( \bm{\mu_1}-\bm{\mu_2}+\frac{1}{2}\bm{r_0}^T\Omega )  )  (\sigma_1+\sigma_2)^{-1}  (-\frac{1}{2}\bm{r_0}^T\sigma _2+i( \bm{\mu_1}-\bm{\mu_2}+\frac{1}{2}\bm{r_0}^T\Omega )  )^T   ]   }\\
    &=    e^{\frac{1}{4}\qty[(\frac{\bm{r_0}^T}{2}(-\sigma_2+i\Omega)+i(\bm{\mu_1}-\bm{\mu_2})  )(\sigma_1+\sigma_2)^{-1}  ((-\sigma_2+i\Omega)^T \frac{\bm{r_0}}{2}+i(\bm{\mu_1}-\bm{\mu_2})^T  )  ]   }\\
    &=e^{\frac{1}{4}\qty[\bm{r_0}^T \frac{(\sigma_2-i\Omega)^T}{2}  (\sigma_1+\sigma_2)^{-1}\frac{(\sigma_2-i\Omega)}{2}\bm{r_0}   ]} e^{\qty(\frac{-i}{2}(\bm{\mu_1}-\bm{\mu_2})^T(\sigma_1+\sigma_2)^{-1}(\frac{\sigma_2-i\Omega}{2})  )\bm{r_0} } e^{-\frac{1}{4}(\bm{\mu_1}-\bm{\mu_2})^T(\sigma_1+\sigma_2)^{-1}  (\bm{\mu_1}-\bm{\mu_2})}.
\end{align*}
Thus, we have for all integrals
\begin{align}
    &\frac{1}{(2\pi)^n}    \frac{1}{\sqrt{\det(\sigma_1+\sigma_2)}}  e^{-\frac{1}{4}(\bm{\mu_1}-\bm{\mu_2})^T(\sigma_1+\sigma_2)^{-1}  (\bm{\mu_1}-\bm{\mu_2})} \nonumber \\
    &\quad \times \int_{\mathds{R}^{2n}} \dd\bm{r_0}\, e^{-\frac{1}{4}\bm{r_0}^T\qty[\sigma_0 + \sigma_2-  \frac{(\sigma_2-i\Omega)^T}{2}(\sigma_1+\sigma_2)^{-1} \frac{(\sigma_2-i\Omega)}{2}]\bm{r_0}  }
    e^{-i\qty[ \bm{\mu_0}-\bm{\mu_2}+\frac{1}{2}(\bm{\mu_1}-\bm{\mu_2}) (\sigma_1+\sigma_2)^{-1}  \frac{(\sigma_2-i\Omega)}{2}  ]\bm{r_0}}\\
    &=\frac{1}{\sqrt{\det(\sigma_1+\sigma_2)}} \frac{1}{\sqrt{\det(\sigma_0+\Delta)}} e^{-\frac{1}{4}(\bm{\mu_1}-\bm{\mu_2})^T(\sigma_1+\sigma_2)^{-1}  (\bm{\mu_1}-\bm{\mu_2})}  e^{-\frac{1}{4}(\bm{\mu_0}-\mu_{\Delta})^T(\sigma_0+\Delta)^{-1}(\bm{\mu_0}-\bm{\mu_{\Delta}}) },
\end{align}
where we take
\begin{align}
    \Delta&=\sigma_2-  \frac{(\sigma_2-i\Omega)^T}{2}(\sigma_1+\sigma_2)^{-1} \frac{(\sigma_2-i\Omega)}{2},\\
    \bm{\mu_{\Delta}}&=\bm{\mu_2}+\frac{1}{2}(\bm{\mu_1}-\bm{\mu_2}) (\sigma_1+\sigma_2)^{-1}  \frac{(\sigma_2-i\Omega)}{2} .
\end{align}
Consequently, we obtain
\begin{align*}
    \bra{G_2}\ket{G_0}\bra{G_1}\ket{G_2}\bra{G_0}\ket{G_1} &= \frac{1}{\sqrt{\det(\sigma_1+\sigma_2)}} \frac{1}{\sqrt{\det(\sigma_0+\Delta)}} e^{-\frac{1}{4}(\bm{\mu_1}-\bm{\mu_2})^T(\sigma_1+\sigma_2)^{-1}  (\bm{\mu_1}-\bm{\mu_2})}  \\
    &\times e^{-\frac{1}{4}(\bm{\mu_0}-\bm{\mu_{\Delta}})^T(\sigma_0+\Delta)^{-1}(\bm{\mu_0}-\bm{\mu_{\Delta}}) }.
\end{align*}

\section{Phases and reference states}
\label{ap:refstate}
As we have shown in Appendix~\ref{ap:innerproduct}, we can compute the inner product between two pure Gaussian states using the covariance matrix, the mean, and a reference state.
Any $n$-mode Gaussian unitary can be decomposed as
\begin{align}
    G=U S(\gamma) D(\beta)V,
\end{align}
where $U,V$ are passive symplectic transformations, $S$ the single-mode squeezing, and $D$ the displacement~\cite{chabaud2022holomorphic}. 
Any pure Gaussian state can be consequently written as 
\begin{align}
    \ket{G}&=US(\gamma) D(\beta)V\ket{0}\\
    &=U \qty(\bigotimes_{i=1}^n S(\gamma_i) D(\beta_i)  )\ket{0}\\
    &= U \bigotimes_{i=1}^n \ket{G_i},
\end{align}
where we use $V\ket{0}=\ket{0}$.
We choose the vacuum $\ket{0}$ as our reference state.
We thus want to compute the overlap 
\begin{align}
    \bra{0}\ket{G}= \bra{0}U \bigotimes_{i=1}^n \ket{G_i}= \bigotimes_{i=1}^n\bra{0}\ket{G_i}.
\end{align}

We can compute this overlap by using the stellar representation~\cite{chabaud2022holomorphic}; in particular with~\cite[Lemma 7]{chabaud2022holomorphic}, we see that
\begin{align}
\label{eq:overlap_0_G}
     \bigotimes_{i=1}^n\bra{0}\ket{G_i} = \prod_{i=1}^n (1-\abs{t_i}^2)^{-\frac{1}{4}} e^{C_i}
\end{align}
with $t_i= e^{i \theta_i } \tanh(r_i) $, $\gamma_i = e^{i \theta_i} r_i$, and $C_i=  \frac{1}{2}(t^*_i \beta_i^2-\abs{\beta_i}^2  )$ for $\gamma_i=r_i e^{i\theta_i}$.

We can retrieve these coefficients from a given covariance matrix and mean.
For this retrieval, our simulator keeps track of the mean and the covariance matrix during the applications of the Gaussian unitary.
The mean and covariance matrix before the application of $U$ are those of single-mode Gaussians
\begin{align}
    \mu &= \bigoplus_{i=1}^n S_i \beta_i,\\
    \sigma&= \bigoplus_{i=1}^n  S_i \mathds{1}_i S_i^T = \bigoplus_{i=1}^n \sigma_i,
\end{align}
where $S_i$ is the symplectic matrix associated to $S(\gamma_i)$.
Then, the application of $U$ yields
\begin{align}
        \mu &= \Bar{U}\bigoplus_{i=1}^n S_i \beta_i,\\
    \sigma&= \Bar{U}\bigoplus_{i=1}^n  \sigma_i  \Bar{U}^T,
\end{align}
where $\Bar{U}$ is the orthogonal symplectic matrix associated with the unitary $U$.
We can thus use $\sigma$ to find $\Bar{U}$ by block-diagonalizing $\sigma$ and consequently $S_i$ from $\sigma_i$.
Having access to $\Bar{U}$ and $S_i$, we can also compute $\beta$ from the mean $\mu$.
Numerical recipes can be found in~\cite{houde2024matrixdecompositionsquantumoptics}.

\section{Alternative phase sensitive simulator}
\label{ap:alt_phase}
In this section, we give an alternative phase-sensitive simulator for Gaussian overlaps.
We base it on~\cite{yao2024riemannianoptimizationphotonicquantum}.
We  use the stellar formalism directly in the computation.
The stellar function of a Gaussian pure state $\dyad{\psi}$ consisting of M modes is given by
\begin{align}
    \Gamma\qty(\bm{\alpha}) =c_{\psi} \exp\qty(\bm{\alpha}^T \bm{b_\psi}+\frac{1}{2}\bm{\alpha}^T A_{\psi}\bm{\alpha}),
\end{align}
where $A_{\psi}$ is an $M\times M$ complex symmetric matrix, $\bm{b_\psi}$ is a $M-$dimensional complex vector , $c_{\psi}$ the vacuum amplitude.
The same can be done for Gaussian mixed states $\rho$ with $A_\rho$, $\bm{b_\rho}$, $c_\rho$ where $A_\rho$, $\bm{b_\rho}$ are $2M$ dimensional.
In this case, we can compute these parameters directly from the covariance matrix and mean of the Gaussian state $\rho$. We will use the $s-$parameterized complex covariance matrix $\bar{\sigma}_S$
\begin{align}
    \bar{\sigma}_S= \bar{\sigma}+\frac{s}{2}\mathds{1}_{2M}
\end{align}
with
\begin{align}
    \bar{\sigma}= \bar{U}^\dagger  \sigma\bar{U}
\end{align}
and the complex mean
\begin{align}
    \bm{\bar{\mu}}=\bar{U}^\dagger\bar{r}.
\end{align}
We use the matrix
\begin{align}
    \bar{U}=\bigoplus_{j=1}^M\frac{1}{\sqrt{2}} \begin{pmatrix}
        1 & i \\
        1 & -i 
    \end{pmatrix}
\end{align}
to go between the quadrature operators $\bm{r}$ and the annihilation and creation operators $a,a^\dagger$.
We will use the matrix $W$ to switch between the ordering of the quadratures operators $r_{q_1},r_{p_1},...,r_{q_M},r_{p_M}$ to $r_{q_1},...,r_{q_M}, r_{p_1},...$.

Then we can compute the matrix $A_\rho$, the vector $\bm{b_\rho}$ and $c_\rho$ using the covariance matrix and the mean as follows:
\begin{align}
    A_\rho &= P_M W \bar{\sigma}_{+1}^{-1} \bar{\sigma}_{-1} W^\dagger,\\
    \bm{b_\rho} &=P_M W \eta_{+1}^{-1}\bm{\bar{\mu}},\\
    c_{\rho}&=\frac{\exp\qty(-\frac{1}{2}\bm{\bar{\mu}}^\dagger \sigma_{+1}^{-1}  \bm{\bar{\mu}})}{\sqrt{\det\qty(\sigma_{+1})}},\\
    P_M&= \begin{pmatrix}
        0_M& \mathds{1}_M\\
        \mathds{1}_M & 0_M
    \end{pmatrix}.
\end{align}

Note that the map $\bar{\sigma}\rightarrow \bar{\sigma}_{+1}^{-1}\bar{\sigma}_{-1}$ is known as the Caley transform and for a pure state $\dyad{\psi}$. It holds that
\begin{align}
    A_{\rho}&= A^*_\psi \oplus A_\psi,\\
    \bm{b_\rho}&=\bm{b_\psi}^*\oplus  \bm{b_\psi}.
\end{align}
So we see that for pure Gaussian states, we only need dimensionality $M$ instead of $2M$.

For a single mode Gaussian state $\ket{\psi}= D(\alpha) S(r e^{i\phi})\ket{0}$ we can give these parameters directly
\begin{align}
    A_\psi&= -\tanh(r)e^{i\phi},\\
    \bm{b_\psi}&=\alpha +\alpha^*e^{i\phi}\tanh(r),\\
    c_\psi&= \frac{\exp\qty(-\frac{1}{2}\qty[\abs{\alpha}^2+\alpha^{*2}e^{i\phi}\tanh(r) ])}{\sqrt{\cosh(r)}}.
\end{align}

If we want to compute the overlap between Gaussian states, we want to compute quantities like
\begin{align}
\label{eq:stellar_overlap}
    \bra{\bm{\alpha}^*} U \ket{\bm{\beta}}=\exp\qty(-\frac{1}{2}\qty[\abs{\bm{\alpha}}^2+\abs{\bm{\beta}}^2]) c_U \exp\qty( \bm{b_U}^T\bm{\nu}+\frac{1}{2} \bm{\nu}^ A_U \bm{\nu})
\end{align}
with $\bm{\nu}=(\bm{\alpha}, \bm{\beta})^T$, where $\ket{\alpha},\ket{\beta}$ are multimode coherent states, and $U$ a Gaussian unitary.
These parameters can be computed directly from the symplectic matrix $S$ and the displacement $\bm{d}$ corresponding to $U$ as follows:
\begin{align}
    A_U&= P_{2M} R \begin{pmatrix}
        \mathds{1}_{2M}-\bm{\xi}^{-1}& \bm{\xi}^{-1} S\\
        S^T \bm{\xi}^{-1}& \mathds{1}_{2M}- S^T \bm{\xi}^{-1}S
    \end{pmatrix}R^\dagger,\\
    \bm{b_U}&= R^* \begin{pmatrix}
        \bm{\xi}^{-1} \bm{d}\\
        -S^T \bm{\xi}^{-1} \bm{d}
    \end{pmatrix},\\
    c_U&= \frac{\exp\qty(-\frac{1}{2} \bm{d}^T \bm{\xi}^{-1} \bm{d})}{\sqrt{\det(\bm{\xi})}},
\end{align}
with 
\begin{align}
    R&=\frac{1}{\sqrt{2}} \begin{pmatrix}
        \mathds{1}_M& i  \mathds{1}_M &0_M &0_M\\
        0_M & 0_M &  \mathds{1}_M & -i  \mathds{1}_M\\
         \mathds{1}_M& -i  \mathds{1}_M&0_M&0_M\\
         0_M&0_M&  \mathds{1}_M&i  \mathds{1}_M
    \end{pmatrix},\\
    \bm{\xi}&=\frac{1}{2}\qty(\mathds{1}_{2M}+SS^T).
\end{align}
This can be computed faster by simplifying it further, see~\cite{yao2024riemannianoptimizationphotonicquantum}.  

If two unitaries are applied one after another $U_f = U_1 U_2$, it is not enough to just consider $S=S_1 S_2$ and $\bm{d}=\bm{d_1}+\bm{d}_2$ as the phase information is lost this way. However, one can update the parameters  $A_{U_f}, \bm{b_{U_f}},c_{U_f}$ to compute the overlap in a phase sensitive way.
For two Gausian unitaries $U_f = U_1 U_2$, the parameters are updates as follows:
\begin{align}
    A_{U_f}&= B_1 \oplus D_2+ \{C_1 \oplus C_2^T  \} \mathcal{Z}\{C_1^T \oplus C_2  \},\\
    \bm{b_{U_f}}^T &= [\bm{c_1}^T, \bm{d_2}^T] [ \bm{d_1}^T, \bm{c_2}^T] \mathcal{Z}\{C_1^T \oplus C_2  \},\\
    c_{U_f}&=\frac{c_{U_1}c_{U_2}}{\sqrt{\det(Y)}} \exp\qty(\frac{1}{2}[\bm{d_1}^T, \bm{c_2}^T]\mathcal{Z}[\bm{d_1},\bm{c_2}]^T),
\end{align}
where we used the notation
\begin{align}
    \bm{b_{U_i}}^T&=[\bm{c_i}^T,\bm{d_i}^T],\\
    A_{U_i}&=\begin{pmatrix}
        B_i &C_i\\
        C_i^T D_i
    \end{pmatrix},
\end{align}
and the auxiliary quantities
\begin{align}
    Y&=\mathds{1}_M-D_1B_2,\\
    \mathcal{Z}&=\begin{pmatrix}
      Y^{-1}B_2& Y^{-1}\\
      (Y^T)^{-1} & D_1 Y^{-1}
    \end{pmatrix}.
\end{align}
If more Gaussian unitaries are applied one just updates the parameters consecutively.
Using Eq.~\eqref{eq:stellar_overlap} one can then compute the overlap using the updated parameters.

\section{Simulation algorithm based on Gaussian extent}
\label{ap:Gextent}

\subsection{Finite rank approximations}

If the given non-Gaussian state $\ket{\psi}$ has an infinite Gaussian rank, such as the Fock state $\ket{1}$,
\begin{align}
    \ket{\psi}=\sum_{i=0}^\infty c_i \ket{G_i},\quad |c_0|\geq|c_1|\geq|c_2|\geq\cdots,
\end{align}
then we consider a state with a cut-off
\begin{align}
    \ket{\Tilde{\psi}_m}&=\sum_{i=0}^m c_i \ket{G_i},\\
    \ket{\psi_m}&=\frac{1}{N_m}\ket{\Tilde{\psi}_m},
\end{align}
with normalization
\begin{equation}
    N_m=\sqrt{\braket{\Tilde{\psi}_m}}=\sqrt{\sum_{i,j=0}^m c_i^*c_j \bra{G_i}\ket{G_j}}.
\end{equation}
We can check that
\begin{align}
    \lim_{m \rightarrow\infty}&\bra{\psi}\ket{\psi_m}=1,\\
    \lim_{m \rightarrow\infty}& N_m=1.
\end{align}
In the following, we will work on this finite-rank approximation.

\subsection{Standard Approach}
This argument is based on Ref.~\cite{bravyi2019simulation}.
We write our input state as 
\begin{align*}
    \ket{\psi}=\sum_{i=1}^\chi c_i \ket{G_i},
\end{align*}
where $\ket{G_i}$ is a Gaussian state characterized by a covariance matrix $\sigma_i$ and a mean $\mu_i$, so we will use the notation $\ket{G_i}=\ket{\sigma_i, \mu_i}$.
We want to estimate the Born probability
\begin{align}
    P(x)&=\frac{\braket{\psi}{x}\braket{x}{\psi} }{\norm{\psi}^2}\\
    &= \frac{\sum_{i,j=1}^{\chi}\braket{\sigma_j, \mu_j}{x}\braket{x}{\sigma_i, \mu_i} }{\norm{\psi}^2}.
\end{align}
Estimating the Born probability takes $\mathcal{O}(\chi^2)$ inner products.
This can be improved by sampling terms $\ket{\sigma_i, \mu_i}$ in the decomposition to sparsify $\ket{\psi}$ according to
\begin{align}
    \ket{\psi}&=\sum_{i=1}^\chi c_i\ket{\sigma_i, \mu_i}\\
    &=\norm{c}_1\sum_{i=1}^\chi \frac{\abs{c_i}}{\norm{c}_1}\frac{c_i}{\abs{c_i}}  \ket{\sigma_i, \mu_i}\\
    &=\norm{c}_1\sum_{i=1}^\chi p(i)\frac{c_i}{\abs{c_i}}  \ket{\sigma_i, \mu_i}\\
    &=\norm{c}_1\sum_{i=1}^\chi p(i) \ket{\Tilde{\sigma}_i, \Tilde{\mu}_i}.
\end{align}

We now consider sampling the states $\ket{\Tilde{\sigma}_i, \Tilde{\mu}_i}$ with probability $p(i)$.
To do this, we define a random variable $\ket{\omega_i}$ that is  $\ket{\Tilde{\sigma}_i, \Tilde{\mu}_i}$ with probability $p(i)$.
By definition, it holds that $\mathds{E}[\ket{\omega_i}]=\frac{\ket{\psi}}{\norm{c}_1}$ and consequently
\begin{align}
    \ket{\psi}=\norm{c}_1 \mathds{E}[\ket{\omega_i}].
\end{align}
Performing this sampling $k$ times, we obtain a sparsified state
\begin{align}
    \ket{\Omega}
    &=\frac{\norm{c}_1}{k}\sum_{i=1}^k \ket{\omega_i}.
\end{align}

By definition, it holds that
\begin{align}
    \mathds{E}\qty[\bra{\Omega} \ket{\psi}] = \mathds{E}[  \bra{\psi }\ket{\Omega}] =1.
\end{align}
The norm of $\ket{\Omega}$ is
\begin{align}
    \mathbb{E}[\bra{\Omega}\ket{\Omega}] &=\frac{\norm{c}_1^2}{k^2}\qty(\mathds{E}\qty[\sum_{\alpha=1}^k\bra{\omega_\alpha}\ket{\omega_\alpha}  ]+ \mathds{E}\qty[ \sum_{\alpha\neq \beta}\bra{\omega_\alpha} \ket{\omega_\beta} ])\\
    &= \frac{\norm{c}_1^2}{k}\mathds{E}[\bra{\omega_\alpha}\ket{\omega_\alpha}] + \frac{1}{k^2}  \sum_{\alpha\neq \beta}\mathds{E}\qty[\norm{c}_1^2 \bra{\omega_\alpha}\ket{\omega_\beta}]\\
    &=  \frac{\norm{c}_1^2}{k} + \frac{1}{k^2}\sum_{\alpha\neq\beta}\bra{\psi}\ket{\psi}\\
    &=  \frac{\norm{c}_1^2}{k} + \frac{k(k-1)}{k^2}\\
    &= \frac{\norm{c}_1^2}{k} + 1-\frac{1}{k}.
\end{align}

The bound of the approximation is
\begin{align}
    &\mathds{E}[\abs{\ket{\psi}-\ket{\Omega}}^2 ]\nonumber\\
    &= \mathds{E}[\bra{\Omega}\ket{\Omega}] +\mathds{E}[\bra{\psi}\ket{\psi}]-\mathds{E}[\bra{\psi}\ket{\Omega}]-\mathds{E}[\bra{\Omega}\ket{\psi}]\\
    &=\frac{\norm{c}_1^2-1}{k} \leq  \frac{\norm{c}_1^2}{k}.
\end{align}
To guarantee the upper bound $\mathds{E}[\|\ket{\psi}-\ket{\Omega}\|^2 ] \leq \delta^2$, it suffices to sample $k$ times with
\begin{equation}
    k=\qty(\frac{\norm{c}_1}{\delta})^2.
\end{equation}

In Ref.~\cite{bravyi2019simulation}, a sparsification tail bound is given.
If we choose  $k\geq \qty(\frac{\norm{c}_1}{\delta})^2$, then we have
\begin{align}
    \mathds{E}[\bra{\Omega}\ket{\Omega}-1]\leq \delta^2.
\end{align}
Using the triangle inequality, we get
\begin{align}
    \norm{\psi-\Omega}^2\leq \bra{\Omega}\ket{\Omega}-1+2\times\abs{1-\text{Re}(\bra{\psi}\ket{\Omega})}.
\end{align}
We define a random variable
\begin{align}
    X_\alpha&=\norm{c}_1 \text{Re}(\bra{\psi}\ket{\omega_\alpha}),\\
    \Tilde{X}&= \frac{1}{k}\sum_{\alpha=1}^k X_\alpha= \text{Re}(\bra{\psi}\ket{\Omega}).
\end{align}
Then, we have
\begin{align}
    \abs{\text{Re}(\bra{\psi}\ket{\Omega})-1} = \abs{\Tilde{X}-\mathds{E}[\Tilde{X}]}
\end{align}
Here, $\Tilde{X}$ is a sample mean of $k$ IID random variables $X_\alpha$, satisfying
\begin{align}
    \abs{X_\alpha}\leq \norm{c}_1 \abs{\bra{\psi}\ket{w_\alpha}} \leq \norm{c}_1 \sqrt{F(\psi)},
\end{align}
with $F$ being the Gaussian fidelity $F(\psi)= \sup_{\ket{G} \in \mathcal{G}}\{\abs{\bra{\psi}\ket{G}}^2\} $.
We can then apply Hoeffding's inequality to obtain
\begin{align}
    &\text{Pr}\qty[\abs{\text{Re}(\bra{\psi}\ket{\Omega})-1}\geq\frac{\delta^2}{2}    ]\nonumber\\
    &\leq 2 \exp\qty(-\frac{2k(\delta^2/2)^2}{(2\norm{c}_1\sqrt{F(\psi)}   )^2}  )\\
    &\leq 2 \exp\qty(-\frac{\delta^2}{8 F(\psi)}).
\end{align}
Consequently, it holds that
\begin{align}
\text{Pr}[\norm{\psi-\Omega}^2\leq \bra{\Omega}\ket{\Omega}-1+\delta^2]\geq 1-2\exp\qty(-\frac{\delta^2}{8F(\psi)}).
\end{align}

\subsection{Approach by Seddon et al.}

We describe another approach based on Ref.~\cite{seddon_quantifying_2021}.
If we use the approach described in the previous section, post-selection is needed for states with a norm close to $1$, but Ref.~\cite{seddon_quantifying_2021} proposes a sampling strategy that avoids post-selection.
This strategy renormalizes and bounds the error between $\ket{\psi}$ and the ensemble 
\begin{align}
    \rho_1 = \mathds{E}\qty(\frac{\dyad{\Omega}}{\braket{\Omega}} ) = \sum_\Omega \text{Pr}(\Omega) \frac{\dyad{\Omega}}{\braket{\Omega}}
\end{align}
of sparsified states.
We then want to bound
\begin{align}
    \norm{\rho_1 -\dyad{\psi}}_1.
\end{align}
First, we define
\begin{align}
    \rho_2= \frac{1}{\mu} \mathds{E}[\dyad{\Omega}].
\end{align}
with $\mathds{E}[\bra{\Omega}\ket{\Omega}]$.
The triangle inequality yields
\begin{align}
\label{eq:triangle_appendix}
    \norm{\rho_1+\rho_2-\rho_2-\dyad{\psi}}_1\leq \norm{\rho_1-\rho_2}_1+\norm{\rho_2-\dyad{\psi}}_1.
\end{align}
The first term on the right-hand side is given by
\begin{align}
    \norm{\rho_1-\rho_2}_1 =\norm{ \mathds{E}\qty[\dyad{\Omega}\qty( \frac{1}{\bra{\Omega}\ket{\Omega}}-\frac{1}{\mu})   ]}_1.
\end{align}
We then use Jensen's inequality to obtain
\begin{align}
    \norm{\rho_1-\rho_2}_1 &\leq \mathds{E}\qty[ \norm{\dyad{\Omega}\qty(  \frac{1}{\bra{\Omega}\ket{\Omega}}-\frac{1}{\mu}  )   }_1 ]\\
    &= \mathds{E}\qty[ \abs{\bra{\Omega}\ket{\Omega}\qty( \frac{1}{\bra{\Omega}\ket{\Omega}}-\frac{1}{\mu})   }]\\
    &= \frac{1}{\mu} \mathds{E}\qty[\abs{\mu-\bra{\Omega}\ket{\Omega}}  ]\\
    &\leq \sqrt{\mathds{E}\abs{\mu -\bra{\Omega}\ket{\Omega} }^2}\\
    &= \sqrt{\text{Var}[\bra{\Omega}\ket{\Omega}]}.
\end{align}
To evaluate the second term on the right-hand side of~\eqref{eq:triangle_appendix}, we use
\begin{align}
    \mu \rho_2 &= \frac{\norm{c}_1^2}{k^2}\qty(\sum_{\alpha\neq \beta }\mathds{E}[\ket{w_\alpha}\bra{w_\beta}  +\sum_\alpha \dyad{\alpha}])\\
    &= \frac{k(k-1)}{k^2}\dyad{\psi}+\frac{\norm{c}_1^2}{k}\sigma
\end{align}
with $\sigma=\dyad{w_\alpha}$.
Then, we have
\begin{align}
    \norm{\rho_2- \dyad{\psi}}_1&=\frac{1}{\mu}\norm{(1-k^{-1}-\mu)\dyad{\psi}+\frac{\norm{c}_1^2}{k}\sigma}_1\\
    &=\frac{\norm{c}_1^2}{k \mu } \norm{\sigma-\dyad{\psi}}\\
    &\leq 2 \frac{\norm{c}_1^2}{k},
\end{align}
where we use $\mu^{-1}\leq 1$, the triangle inequality, and $\norm{\sigma}_1=1$.

As for the bounds of the variance, we have
\begin{align}
    \text{Var}(\bra{\Omega}\ket{\Omega})=\mathds{E}[\bra{\Omega}\ket{\Omega}^2] -\mathds{E}[\bra{\Omega}\ket{\Omega}]^2.
\end{align}
The right-hand side is evaluated by
\begin{align}
    \bra{\Omega}\ket{\Omega}^2=\frac{\norm{c}_1^4}{k^4}\qty(k^2+ 2 k B +B^2) 
\end{align}
with $B=\sum_\alpha \sum_{\alpha\neq \beta }\bra{w_\alpha}\ket{w_\beta} $.
Thus, we obtain
\begin{align}
    \mathds{E}[\bra{\Omega}\ket{\Omega}^2] =\frac{\norm{c}_1^4}{k^4}\qty(k^2+ 2 k \mathds{E}[B] +\mathds{E}[B^2]) .
\end{align}
Therefore, we have
\begin{align}
      \text{Var}(\bra{\Omega}\ket{\Omega}) = \frac{\norm{c}_1^4}{k^4}\qty(\mathds{E}[B^2] -\mathds{E}[B]^2   ).
\end{align}
Following the same calculation for matching the coefficients as that found in the appendix of Ref.~\cite{seddon_quantifying_2021}, we arrive at
\begin{align}
    \text{Var}(\bra{\Omega}\ket{\Omega}) \leq \frac{4 (C-1)}{k}+2 \qty(\frac{\norm{c}_1^2}{k})^2+\mathcal{O}\qty(\frac{C}{k^3})
\end{align}
with $C=\norm{c}_1 \sum_i \abs{c}_i \abs{\bra{\psi}\ket{\Tilde{\sigma}_i,\Tilde{\mu}_i}}^2$.

Then, there is a critical precision $\delta_c=8 (C-1)/\norm{c}_1^2$ such that for every target precision $\delta_S$ for which $\delta_c\leq \delta_S$,
we can sample pure states from an ensemble $\rho_1$, where
every pure state drawn from $\rho_1$ has Gaussian rank at most $\lceil 4 \norm{c}_1^2/\delta_S  \rceil$.
This bound provides a critical threshold, where we get an improvement by a factor of $1/\delta_S$. For higher precision, there is no improvement obtained from this sampling strategy outside of avoiding post-selection.

\section{Fast norm estimation}
\label{ap:fastnorm}
\subsection{Description of the core idea}
A naive estimation of the norm of a decomposed non-Gaussian state would require a quadratic cost of the number of Gaussian states in the decomposition. 
To achieve better efficiency, we can estimate the norm probabilistically by sampling random coherent states and computing the overlap.
This procedure, called the fast norm estimation, results in a linear cost of the number of Gaussian states in the decomposition.
To achieve this scaling, we sample a coherent state $\ket{\alpha}$ and use the completeness condition
\begin{align}
    \mathds{1}= \frac{1}{\pi} \int_{\mathds{C}} \dd \alpha\, \dyad{\alpha}.
\end{align}
A non-trivial part of the analysis arises from the fact that the right-hand side of this completeness condition uses the integral over all $\alpha$, but it is not straightforward to sample $\alpha$ from a uniform distribution since the displacement operators are a non-compact group and thus impossible to uniformly sample.
To address this point, we need to sample $\alpha$ from a weighted distribution in reality.
Nevertheless, in this section, we use the integral without weight to describe the core idea of our fast norm estimation procedure, while we will explain the sampling from the weighted distribution in the next section.

Using the uniform integral of a random coherent state $\ket{\bm{\xi}} = D(\bm{\xi})\ket{0}$, we can define a variable $X$ and $\ket{\psi}$
\begin{align}
    X=\pi^{-n} \abs{ \bra{\bm{\xi}}\ket{\psi}}^2.
\end{align}
In this case, we have
\begin{equation}
 \int_\mathbb{C}\dd \bm{\xi} X= \frac{1}{\pi} \int_\mathbb{C}d\alpha\braket{\psi}{\alpha}\braket{\alpha}{\psi} = \braket{\psi} .   
\end{equation}

Similarly, we can write for $X^2$ as
\begin{align}
    \int_{\mathds{C}^n} \dd \bm{\xi} X^2 = \pi^{-2n}   \int_{\mathds{C}^n} \dd \bm{\xi}\bra{\psi\otimes \psi}  D(\bm{\xi})\otimes D(\bm{\xi}) \dyad{0 \otimes 0}D^\dagger(\bm{\xi}) \otimes D^\dagger(\bm{\xi})    \ket{\psi\otimes\psi}.
\end{align}
So we want to investigate
\begin{align}
\label{eq:def_T}
   \mathcal{T}(\dyad{0\otimes 0})\coloneqq \int_{\mathds{C}^n} \dd \bm{\xi}\,  D(\bm{\xi})\otimes D(\bm{\xi}) \dyad{0 \otimes 0}D^\dagger(\bm{\xi}) \otimes D^\dagger(\bm{\xi}) 
\end{align}
in order to get further insights into $X^2$.
It holds in general for displacement operators that
\begin{align}
    &D(\bm{\xi})D(\bm{\alpha})D(\bm{\xi})^\dagger\nonumber\\
    &=  D(\bm{\xi})D(\bm{\alpha})D(-\bm{\xi})\\
    &=  e^{\frac{1}{2}\qty( \bm{\xi} \bm{\alpha}^*-\bm{\xi}^* \bm{\alpha}   )} D(\bm{\xi}+\bm{\alpha}) D(-\bm{\xi})\\
    &=e^{\frac{1}{2}\qty( \bm{\xi} \bm{\alpha}^*-\bm{\xi}^* \bm{\alpha}   )}  e^{\frac{1}{2}\qty( (\bm{\xi}+\bm{\alpha})(-\bm{\xi})^*-(\bm{\xi}+\bm{\alpha})^*(-\bm{\xi})    )}  D(\bm{\alpha})\\
    &=e^{\frac{1}{2}\qty( \bm{\xi} \bm{\alpha}^*-\bm{\xi}^* \bm{\alpha}   )}  e^{\frac{1}{2}\qty( -\bm{\xi}\bm{\xi}^*-\bm{\alpha}\bm{\xi}^*+\bm{\xi}^*\bm{\xi}+\bm{\alpha}^*\bm{\xi}      )}  D(\bm{\alpha})\\
    &= e^{ \bm{\xi} \bm{\alpha}^*-\bm{\xi}^* \bm{\alpha}   }D(\bm{\alpha}).
\end{align}
We can expand every state in terms of the basis of displacement operators with the characteristic function $\chi_\rho$ as
\begin{align}
    \rho =\frac{1}{\pi^{2n}} \int_{\mathds{C}^n} \dd \bm{\alpha} \dd \bm{\beta}\; \chi_{\rho}(\bm{\alpha},\bm{\beta})  D(\bm{\alpha})\otimes D(\bm{\beta}).
\end{align}
So $\mathcal{T}(\dyad{0\otimes 0})$ can be computed using the following property
\begin{align}
    &\int_{\mathds{C}^n} \dd\bm{\xi}\, \left[D(\bm{\xi})\otimes D(\bm{\xi})\right] \left[D(\bm{\alpha})\otimes D(\bm{\beta})\right] \left[D^\dagger (\bm{\xi})\otimes D^\dagger (\bm{\xi})\right]\\
    &=\int_{\mathds{C}^n} \dd\bm{\xi}\, ( D(\bm{\xi})D(\bm{\alpha})D(\bm{\xi})^\dagger)\otimes  ( D(\bm{\xi})D(\bm{\beta})D(\bm{\xi})^\dagger)\\
    &= \int_{\mathds{C}^n} \dd\bm{\xi}\,  e^{ \bm{\xi} \bm{\alpha}^*-\bm{\xi}^* \bm{\alpha}   } e^{ \bm{\xi} \bm{\beta}^*-\bm{\xi}^* \bm{\beta}   }   D(\bm{\alpha})\otimes D(\bm{\beta})\\
    &= \int_{\mathds{C}^n} \dd\bm{\xi}\,  e^{ \bm{\xi} (\bm{\alpha}^*+\bm{\beta}^*)-\bm{\xi}^* (\bm{\alpha}+\bm{\beta})   }   D(\bm{\alpha})\otimes D(\bm{\beta})\\
    &= \pi^n  \delta (\bm{\alpha}+\bm{\beta}) D(\bm{\alpha})\otimes D(\bm{\beta})
\end{align}
and then
\begin{align}
    \mathcal{T}(\dyad{0\otimes 0}) &=   \pi^{-2n}  \int_{\mathds{C}^n} \dd \bm{\alpha}\; \dd \bm{\beta}\; \chi_{\ket{0\otimes 0}}(\bm{\alpha},\bm{\beta}) \mathcal{T}(D(\bm{\alpha})\otimes D(\bm{\beta})) \\
    &=\pi^{-n}\int_{\mathds{C}^n} \dd \bm{\alpha}\; \chi_{\ket{0\otimes 0}}(\bm{\alpha},-\bm{\alpha})D(\bm{\alpha})\otimes D(-\bm{\alpha}).
\end{align}
We will follow the reasoning of Ref.~\cite{Bravyi2017complexity} and show that
that $\mathcal{T}(\dyad{0\otimes 0})$ is proportional to a projector. This will allow us to bound $\bra{\psi \otimes \psi } \mathcal{T}(\dyad{0\otimes 0}) \ket{\psi \otimes \psi }$ from above. So we need to show that $\mathcal{T}(\dyad{0\otimes 0})=\mathcal{T}(\dyad{0\otimes 0})^\dagger$ and $\mathcal{T}(\dyad{0\otimes 0})\propto \mathcal{T}(\dyad{0\otimes 0})^2$.
By definition of $\mathcal{T}$ in Eq.~\eqref{eq:def_T}, we have that
\begin{align}
    \mathcal{T}(\dyad{0\otimes 0})=\mathcal{T}(\dyad{0\otimes 0})^\dagger.
\end{align}
So we just need to show that $\mathcal{T}(\dyad{0\otimes 0})= \mathcal{T}(\dyad{0\otimes 0})^2$.

Thus, it follows that
\begin{align}
 &\mathcal{T}(\dyad{0\otimes 0})  \mathcal{T}(\dyad{0\otimes 0}) =  \pi^{-4n}  \int_{\mathds{C}^n} \dd \bm{\alpha}\; \dd \bm{\beta}\; \dd \bm{c}\; \dd d\; \chi_{\ket{0\otimes 0}}(\bm{\alpha},\bm{\beta}) \chi_{\ket{0\otimes 0}}(\bm{c},d) \\
 &\times \mathcal{T}(D(\bm{\alpha})\otimes D(\bm{\beta}))  \mathcal{T}(D(\bm{c})\otimes D(d)) \\
 &=\pi^{-2n}\int_{\mathds{C}^n} \dd \bm{\alpha}\; \dd \bm{c}\; \chi_{\ket{0\otimes 0}}(\bm{\alpha},-\bm{\alpha}) \chi_{\ket{0\otimes 0}}(\bm{c},-\bm{c}) \qty[D(\bm{\alpha})\otimes D(-\bm{\alpha}) ] \qty[D(\bm{c})\otimes D(-\bm{c}) ]\\
 &=\pi^{-2n}\int_{\mathds{C}^n} \dd \bm{\alpha}\; \dd \bm{c} \;\chi_{\ket{0\otimes 0}}(\bm{\alpha}) \chi_{\ket{0\otimes 0}}(\bm{c}) e^{\bm{\alpha}c^*-\bm{\alpha}^*\bm{c}/2} D(\bm{\alpha}+\bm{c})\otimes \chi_{\ket{0\otimes 0}}(-\bm{\alpha}) \chi_{\ket{0\otimes 0}}(-\bm{c}) e^{\bm{\alpha}\bm{c}^*-\bm{\alpha}^*\bm{c}/2} D(-\bm{\alpha}-\bm{c})\\
 &=\pi^{-2n}\int_{\mathds{C}^n} \dd \bm{\alpha} \dd \bm{\mu}\, \chi_{\ket{0\otimes 0}}(\bm{\alpha}) \chi_{\ket{0\otimes 0}}(\bm{\mu}-\bm{\alpha}) e^{\bm{\alpha}\bm{\mu}^*-\bm{\alpha}^*\bm{\mu}/2} D(\bm{\mu})\otimes \chi_{\ket{0\otimes 0}}(-\bm{\alpha}) \chi_{\ket{0\otimes 0}}(-\bm{\mu}+\bm{\alpha}) e^{\bm{\alpha}\bm{\mu}^*-\bm{\alpha}^*\bm{\mu}/2} D(-\bm{\mu})\\
 &=\pi^{-2n}\int_{\mathds{C}^n} \dd \bm{\mu}\,  D(\bm{\mu})\otimes D(-\bm{\mu})\qty[\int_{\mathds{C}^n} \dd \bm{\alpha}\,  \chi_{\ket{0\otimes 0}}(\bm{\alpha})   \chi_{\ket{0\otimes 0}}(-\bm{\alpha})\chi_{\ket{0\otimes 0}}(\bm{\mu}-\bm{\alpha}) \chi_{\ket{0\otimes 0}}(-\bm{\mu}+\bm{\alpha}) e^{\bm{\alpha}\bm{\mu}^*-\bm{\alpha}^*\bm{\mu}}  ]\\
  &=\pi^{-2n}\int_{\mathds{C}^n} \dd \bm{\mu}\,  D(\bm{\mu})\otimes D(-\bm{\mu})\qty[\int_{\mathds{C}^n} \dd \bm{\alpha} \, \abs{\chi_{\ket{0\otimes 0}}(\bm{\alpha})}^2   \abs{\chi_{\ket{0\otimes 0}}(\bm{\mu}-\bm{\alpha})}^2 e^{\bm{\alpha}\bm{\mu}^*-\bm{\alpha}^*\bm{\mu}}  ]\\
    &=\pi^{-n} 2^{-n}\int_{\mathds{C}^n} \dd \bm{\mu}\,  D(\bm{\mu})\otimes D(-\bm{\mu})  \abs{\chi_{\ket{0\otimes 0}}(\bm{\mu})}^2 \\
    &=\pi^{-n} 2^{-n}\int_{\mathds{C}^n} \dd \bm{\mu}\, \chi_{\ket{0\otimes 0}}(\bm{\mu}) \chi_{\ket{0\otimes 0}}(-\bm{\mu}) D(\bm{\mu})\otimes D(-\bm{\mu}).
\end{align}
We used that
\begin{align}
    &\int_{\mathds{C}^n} \dd \bm{\alpha}\,  \abs{\chi_{\ket{0\otimes 0}}(\bm{\alpha})}^2   \abs{\chi_{\ket{0\otimes 0}}(\bm{\mu}-\bm{\alpha})}^2 e^{(\bm{\alpha}m^*-\bm{\alpha}^*\bm{\mu})}  \\
    &=\int_{\mathds{C}^n} \dd \bm{\alpha}\, e^{-\abs{\bm{\alpha}}^2} e^{-\abs{\bm{\mu}-\bm{\alpha}}^2}e^{(\bm{\alpha}\bm{\mu}^*-\bm{\alpha}^*\bm{\mu})} \\
    &=e^{-\abs{\bm{\mu}}^2} \int_{\mathds{C}^n} \dd \bm{\alpha}\, e^{-2\abs{\bm{\alpha}}^2+2\bm{\mu}^* \bm{\alpha}}\\
    &= e^{-\abs{\bm{\mu}}^2}\int_{\mathds{R}^n} \dd \bm{\alpha}_r\, \dd \bm{\alpha}_i\, e^{-2\qty(\bm{\alpha}_r^2+\bm{\alpha}_i^2+\bm{\alpha}_r \bm{\mu}_r-\bm{\alpha}_i \bm{\mu}_i+i \bm{\alpha}_r \bm{\mu}_i -i \bm{\mu}_r \bm{\alpha}_i)}\\
    &=  e^{-\abs{\bm{\mu}}^2}\int_{\mathds{R}^n} \dd \bm{\alpha}_r\, e^{-2 \bm{\alpha}_r^2+\bm{\alpha}_r(2\bm{\mu}_r-2i\bm{\mu}_i) }  \int_{\mathds{R}^n}  \dd \bm{\alpha}_i\,  e^{-2 \bm{\alpha}_i^2+\bm{\alpha}_i(2\bm{\mu}_i+2i\bm{\mu}_r) } \\
    &=e^{-\abs{\bm{\mu}}^2}  \frac{\pi^n}{2^n} e^{(\bm{\mu}_r-i \bm{\mu}_i)^2/2} e^{(\bm{\mu}_i+i \bm{\mu}_r)^2/2}\\
    &= \abs{\chi_{\ket{0\otimes 0}}(\bm{\mu})}^2 \frac{\pi^n}{2^n}
\end{align}
and $\chi_{\ket{0\otimes 0}}(a)=e^{-\frac{1}{2}\abs{a}^2}$.
By comparing the results of $\mathcal{T}(\dyad{0\otimes 0})  \mathcal{T}(\dyad{0\otimes 0})$ with 
\begin{align}
    \mathcal{T}(\dyad{0\otimes 0}) &= \pi^{-n}\int_{\mathds{C}^n} \dd \bm{\alpha}\; \chi_{\ket{0\otimes 0}}(\bm{\alpha},-\bm{\alpha})D(\bm{\alpha})\otimes D(-\bm{\alpha}),
\end{align}
we observe that they are the same up to a scaling factor
\begin{align}
    \mathcal{T}(\dyad{0\otimes 0})^2=2^{-n}\mathcal{T}(\dyad{0\otimes 0}).
\end{align}
By multiplying $\mathcal{T}(\dyad{0\otimes 0})$ with $2^n$, we get the projector $\Pi = 2^n \mathcal{T}(\dyad{0 \otimes 0})$. 
This approach is, however, not realistic.
The displacement operators are a non-compact group and thus impossible to sample uniformly.
We address this problem in the next section.

\subsection{Fast norm estimation by sampling from Gaussian ensemble}
We can make it possible to sample from a distribution of displacements by introducing a Gaussian ensemble of displacements~\cite{zhuang2019scrambling}.
The group of all displacements is not compact, and we cannot sample it uniformly.
Reference~\cite{zhuang2019scrambling} uses sampling from a Gaussian ensemble
\begin{align}
    \mathcal{D}_N=\qty{ D(\bm{\xi}):\bm{\xi} \sim P_D^G(\bm{\xi},N)=\frac{e^{-\abs{\bm{\xi}}^2/N}}{\pi^n N^n}   },
\end{align}
which can be used in CV state tomography and reproduces the identity in the limit
\begin{align}
    \lim_{N\rightarrow\infty}N^n\int_{\mathds{C}^n} \dd\bm{\xi}\,  P_D^G(\bm{\xi},N) D(\bm{\xi})\dyad{0} D^\dagger(\bm{\xi})=\mathds{1}.
\end{align}
It also holds that 
\begin{align}
&\lim_{N\rightarrow\infty}N^n\int_{\mathds{C}^n} \dd\bm{\xi}\,   P_D^G(\bm{\xi},N) D(\bm{\xi})^{\otimes 2}\dyad{0\otimes 0} D^\dagger(\bm{\xi})^{\otimes 2}\nonumber\\
&\quad = \frac{1}{\pi^n}\int_{\mathds{C}^n} \dd\bm{\xi} D(\bm{\xi})^{\otimes 2}\dyad{0\otimes 0} D^\dagger(\bm{\xi})^{\otimes 2}.
\end{align}
Our contribution here is to propose sampling from the Gaussian ensemble as an approximation of the procedure in the previous section. We present an analysis in the following.

The above limits ensure that for any $\delta>0$, there is a sufficiently large $N_\delta$ such that, for a state $\ket{\Omega}$ of interest,
\begin{align}
\label{eq:delta_approx_1}
&\left|N_\delta^n \int_{\mathds{C}^n} \dd \bm{\xi} P_D^G(\bm{\xi},N_\delta) \bra{\Omega}D(\bm{\xi}) \dyad{0} D^\dagger(\bm{\xi})\ket{\Omega} - \braket{\Omega}\right|\leq \delta\braket{\Omega}
\end{align}
and
\begin{align}
\label{eq:delta_approx_2}
    &\Big|\pi^n N_\delta^{n}  \int_{\mathds{C}^n} \dd \bm{\xi} P_D^G(\bm{\xi},N_\delta) \bra{\Omega \otimes \Omega} D(\bm{\xi})\otimes D(\bm{\xi}) \dyad{0\otimes 0} D^\dagger(\bm{\xi})\otimes D^\dagger(\bm{\xi})\ket{\Omega \otimes \Omega}\nonumber\\
    &\quad-\int_{\mathds{C}^n} \dd\bm{\xi} \bra{\Omega \otimes \Omega}(D(\bm{\xi})\otimes D(\bm{\xi}))\dyad{0\otimes 0} (D^\dagger(\bm{\xi})\otimes D^\dagger(\bm{\xi}))\ket{\Omega \otimes \Omega}\Big|\leq \delta\braket{\Omega \otimes \Omega}.
\end{align}

Equivalently to the previous section, using Eq.~\eqref{eq:delta_approx_1}, we define
\begin{align}
    \mathcal{T}_G(\dyad{0\otimes 0}) &\coloneqq \pi^n N_\delta^n\int_{\mathds{C}^n} \dd \bm{\xi}\, \frac{e^{-\abs{\bm{\xi}}^2/N_{\delta}}}{\pi^n N_{\delta}^n} D(\bm{\xi})\otimes D(\bm{\xi}) \dyad{0\otimes 0} D^\dagger(\bm{\xi})\otimes D^\dagger(\bm{\xi})\\
      &=\frac{\pi^n N_\delta^n}{\pi^{2n}} \int_{\mathds{C}^n} \dd \bm{\alpha} \;\dd \bm{\beta}\; \dd \bm{\xi}\, \chi_{\ket{0\otimes 0}}(\bm{\alpha},\bm{\beta})  \frac{e^{-\abs{\bm{\xi}}^2/N_{\delta}}}{\pi^n N_{\delta}^n} (D(\bm{\xi})\otimes D(\bm{\xi})) (D(\bm{\alpha} )\otimes D(\bm{\beta})) (D^\dagger(\bm{\xi})\otimes D^\dagger(\bm{\xi}))\\
        &=\frac{\pi^n N_\delta^n}{\pi^{2n}} \int_{\mathds{C}^n} \dd \bm{\alpha} \;\dd \bm{\beta}\; \dd \bm{\xi}\,\chi_{\ket{0\otimes0}}(\bm{\alpha},\bm{\beta})  \frac{e^{-\abs{\bm{\xi}}^2/N_{\delta}}}{\pi^n N_{\delta}^n} e^{\bm{\xi}(\bm{\alpha}^*+\bm{\beta}^*)-\bm{\xi}^*(\bm{\alpha}+\bm{\beta})} D(\bm{\alpha} )\otimes D(\bm{\beta}) \\
        &=\frac{\pi^n N_\delta^n}{\pi^{2n}} \int_{\mathds{C}^n} \dd \bm{\alpha} \;\dd \bm{\beta}\; \chi_{\ket{0\otimes0}}(\bm{\alpha},\bm{\beta})D(\bm{\alpha} )\otimes D(\bm{\beta})\\
        & \quad\times \int_{\mathds{C}^n} \dd \bm{\xi}_r \; \dd \bm{\xi}_i\,  \frac{e^{-\abs{\bm{\xi}_r}^2/N_{\delta}} e^{-\abs{\bm{\xi}_i}^2/N_{\delta}}}{\pi^n N_{\delta}^n} e^{\bm{\xi}_r(\bm{\alpha}^*+\bm{\beta}^*)-\bm{\xi}_r(\bm{\alpha}+\bm{\beta})}  e^{i \bm{\xi}_i(\bm{\alpha}^*+\bm{\beta}^*)+i\bm{\xi}(\bm{\alpha}+\bm{\beta})}  \\
    &=\frac{\pi^n N_\delta^n}{\pi^{2n}}\int_{\mathds{C}^n} \dd \bm{\alpha} \;\dd \bm{\beta}\; \chi_{\ket{0\otimes0}}(\bm{\alpha},\bm{\beta}) e^{-N_\delta \abs{\bm{\alpha}+\bm{\beta}}^2}     D(\bm{\alpha} )\otimes D(\bm{\beta}),
\end{align}
which will approach the result in the previous section in the limit of $N_\delta\rightarrow \infty$.

Importantly, we can use the results from the previous section that
\begin{align}
   \mathcal{T}(\dyad{0\otimes 0})= \int_{\mathds{C}^n} \dd \bm{\xi}\,  D(\bm{\xi})\otimes D(\bm{\xi}) \dyad{0 \otimes 0}D^\dagger(\bm{\xi}) \otimes D^\dagger(\bm{\xi}) 
\end{align}
is proportional to a projector $\Pi = 2^n\mathcal{T}(\dyad{0\otimes 0})$.
In particular, Eq.~\eqref{eq:delta_approx_2} immediately implies that
\begin{align}
\label{eq:delta_approx_3}
\abs{\pi^n N_\delta^n  \int_{\mathds{C}^n} \dd\bm{\xi}\,   P_D^G(\bm{\xi},N_\delta) \bra{\Omega \otimes \Omega} D(\bm{\xi})^{\otimes 2}\dyad{0\otimes 0} D^\dagger(\bm{\xi})^{\otimes 2}  \ket{\Omega \otimes \Omega}- 2^{-n} \bra{\Omega \otimes \Omega} \Pi \ket{\Omega \otimes \Omega}}\\
 \leq \delta \braket{\Omega \otimes \Omega}.
\end{align}

We can then define a random variable $X$ as
\begin{align}
    X=N_\delta^n \abs{\bra{\bm{\xi}}\ket{\Omega}}^2.
\end{align}
Thus, by sampling coherent states from the Gaussian ensemble, we can resolve the identity and estimate the norm
\begin{align}
    \left|\mathds{E}[X]-\braket{\Omega}\right|=\left|N_\delta^n \int_{\mathds{C}^n} \dd \bm{\xi}\; \frac{e^{-\abs{\bm{\xi}}^2/N_\delta}}{\pi^n N_\delta^n}\bra{\Omega} D(\bm{\xi})\dyad{0} D^\dagger(\bm{\xi}) \ket{\Omega}-\braket{\Omega}\right|\leq\delta\braket{\Omega},
\end{align}
where the inequality follows from Eq.~\eqref{eq:delta_approx_1}.

Due to Eq.~\eqref{eq:delta_approx_3}, the variance can be bounded by
\begin{align}
    \text{Var}[X]&\leq \mathds{E}[X^2]\\
    &= N_\delta^{2n}  \int_{\mathds{C}^n} \dd \bm{\xi} \frac{e^{-\abs{\bm{\xi}}^2/N_\delta}}{\pi^n N_\delta^n} \bra{\Omega \otimes \Omega} D(\bm{\xi})\otimes D(\bm{\xi}) \dyad{0\otimes 0} D^\dagger(\bm{\xi})\otimes D^\dagger(\bm{\xi})\ket{\Omega \otimes \Omega}\\
    &\leq \frac{N_\delta^{n} } {\pi^{n}} 2^{-n} \bra{\Omega \otimes \Omega} \Pi \ket{\Omega \otimes \Omega}+ \delta \braket{\Omega\otimes \Omega}\\
    &\leq \frac{2^{-n}N_\delta^{n}+\delta\pi^n}{\pi^{n}} \norm{\Omega}^4.
\end{align}

Now, if we define an estimator as
\begin{align}
    \eta = \frac{1}{L}\sum_{i=1}^L N_\delta^n \abs{ \bra{\bm{\xi}}\ket{\Omega}}^2.
\end{align}
Then, $\eta$ has expectation value $\norm{\Omega}^2$ up to $\pm\delta$ and variance
$\sigma^2 \leq L^{-1}(2^{-n}N_\delta^{n}+\delta \pi^n)/\pi^n \norm{\Omega}^4$.
Using Chebyshev's inequality and choosing the number of samples as
\begin{equation}
    L=\frac{2^{-n}N_\delta^{n}+\delta \pi^n}{\pi^n}  \epsilon^{-2} p_f^{-1},
\end{equation}
we have that, with a probability of at least $1-p_f$,
\begin{align}
    (1-\epsilon-\delta)\norm{\ket{\Omega}}^2 \leq \eta \leq (1+\epsilon+\delta)\norm{\ket{\Omega}}^2.
\end{align}

\section{Refined Bounds}
\label{ap:refined}
The output of the channel
\begin{equation}
\begin{aligned}
  &\int_{\mathds{C}^n} \dd \bm{\xi} P_D^G(\bm{\xi},M) D(\bm{\xi}) \dyad{0} D^\dagger(\bm{\xi})=\rho^{th}_M
  \end{aligned}\end{equation}
is the thermal state with mean photon number M
\begin{align}
    \rho^{th}_M= \frac{1}{M+1} \sum_{n=0}^\infty \qty(\frac{M}{M+1})^n\dyad{n}.
\end{align}
This channel is known in the literature as the classical-noise channel~\cite{PhysRevA.70.032315}.
Therefore, the integral is upper bounded by
\begin{equation}
\begin{aligned}
  &\int_{\mathds{C}^n} \dd \bm{\xi} P_D^G(\bm{\xi},M) \bra{\Omega}D(\bm{\xi}) \dyad{0} D^\dagger(\bm{\xi})\ket{\Omega}\leq \braket{\Omega}.
  \end{aligned}\end{equation}
Furthermore it holds for $\braket{\Omega}=\sum_{n=0}^\infty \abs{\Omega_n}^2$. Thus 
\begin{align}
    M \bra{\Omega} \rho^{th}_M \ket{\Omega}= \frac{M}{M+1} \sum_{n=0}^\infty \abs{\Omega_n}^2 \qty(\frac{M}{M+1})^n \leq \sum_{n=0}^\infty \abs{\Omega_n}^2 = \braket{\Omega}.
\end{align}

We can lower bound this integral using the mean photon number of the state $\ket{\Omega}$ $ N_\Omega=\frac{\bra{\Omega}n\ket{\Omega}}{\braket{\Omega}} =\frac{1}{\pi^n \braket{\Omega}} \int_{\mathds{C}^n} \dd \bm{\xi} \abs{\xi}^2 \abs{\bra{\xi}\ket{\Omega}}^2$.
It holds for a Gaussian function that $e^{-x^2}\geq (1-x^2)$. So by expanding $P_D^G(\bm{\xi},N)$ to the first order we get
\begin{equation}
\begin{aligned}
  &\int_{\mathds{C}^n} \dd \bm{\xi} P_D^G(\bm{\xi},N) \bra{\Omega}D(\bm{\xi}) \dyad{0} D^\dagger(\bm{\xi}\ket{\Omega}= \int_{\mathds{C}^n} \dd \bm{\xi} \frac{e^{-\abs{\bm{\xi}}^2/N}}{\pi^n N^n}  \abs{\bra{\xi}\ket{\Omega}}^2\\ 
  &\geq \frac{1}{\pi^n N^n} \int_{\mathds{C}^n} \dd \bm{\xi} \qty(1- \frac{\abs{\xi}^2}{N}) \abs{\bra{\xi}\ket{\Omega}}^2\\
  &= \frac{1}{N^n} \braket{\Omega} \qty(1- \frac{N_\Omega}{N}).
  \end{aligned}\end{equation}

We can then define a random variable $X$ as
\begin{align}
    X=N^n \abs{\bra{\bm{\xi}}\ket{\Omega}}^2.
\end{align}
We can then estimate the norm of $\ket{\Omega}$ by sampling coherent states from the Gaussian ensemble, since
\begin{align}
    \mathds{E}(X)= N^n \int_{\mathds{C}^n} \dd \bm{\xi} P_D^G(\bm{\xi},N) \abs{\bra{\bm{\xi}}\ket{\Omega}}^2
\end{align}
with 
\begin{align}
    \braket{\Omega}\qty(1-\frac{N_\Omega}{N})\leq\mathds{E}(X)\leq \braket{\Omega}.
\end{align}

The variance is bounded in the following way.
\begin{align}
    \text{Var}[X]&\leq \mathds{E}[X^2]\\
    &= N^{2n}  \int_{\mathds{C}^n} \dd \bm{\xi} \frac{e^{-\abs{\bm{\xi}}^2/N}}{\pi^n N^n} \bra{\Omega \otimes \Omega} D(\bm{\xi})\otimes D(\bm{\xi}) \dyad{0\otimes 0} D^\dagger(\bm{\xi})\otimes D^\dagger(\bm{\xi})\ket{\Omega \otimes \Omega}\\
    & \leq \frac{N^n}{\pi^n}  \int_{\mathds{C}^n} \dd \bm{\xi} \abs{\bra{\bm{\xi} \otimes\bm{\xi} }\ket{\Omega \otimes \Omega}}^2\\
    &= \frac{N^{n} } {\pi^{n}} 2^{-n} \bra{\Omega \otimes \Omega} \Pi \ket{\Omega \otimes \Omega}\\
    &\leq\frac{N^{n} } { 2^n \pi^{n}}  \norm{\Omega}^4.
\end{align}

\section{Proof of Eq.~\eqref{eq:witness decomposition}}
\label{ap:witness decomposition}

Let $f\subset \mathcal{H}$ be a subset of separable Hilbert space $\mathcal{H}$.
Consider the extent measure with respect to the set $f$ defined by
\bal
 \xi_f (\ket{\psi})& = \inf\lset \left(\sum_i c_i\right)^2  \sbar  \ket{\psi} = \sum_i c_i \ket{\phi_i}, \ket{\phi_i}\in f\rset.
\eal
We also define 
\bal
 \bar{\xi}_f(\ket{\psi}) \coloneqq \inf\lset \mu^2 \sbar \ket{\psi}\in\mu \cl\conv f\rset.  
\eal
This allows for an alternative expression~\cite{Arveson2009maximal,Lami2021framework}
\bal
 \bar{\xi}_f(\ket{\psi}) = \sup \lset |\braket{w}{\psi}|^2 \sbar |\braket{w}{\phi}|\leq 1,\ \forall \ket{\phi}\in f\rset.
 \label{eq:extent closure dual}
\eal
As shown in Ref.~\cite{Lami2021framework}, this quantity coincides with the lower semicontinuous robustness 
\bal
 \bar{\xi}_f(\ket{\psi}) = \underline{R}_\mF(\dyad{\psi})
 \label{eq:extent closure equal robustness}
\eal
for $\mF=\cl\conv\lset \dyad{\phi} \sbar \ket{\phi}\in f\rset$. 

Let us now assume that a decomposition $\ket{\psi}=\sum_i \tilde c_i \ket{\tilde \phi_i}$ with $\ket{\tilde \phi_i}\in f$ satisfies 
\bal
\left(\sum_i |\tilde c_i|\right)^2 = \xi_f(\ket{\psi}) = \bar{\xi}_f(\ket{\psi}) = \underline{R}_\mF(\dyad{\psi}) = \Tr(\tilde W \dyad{\psi})
\label{eq:assumption optimum achieved}
\eal
for an optimal witness operator $\tilde W$, which appears in the definition of $\underline{R}_\mF$. 
Note that
\bal
 \underline{R}_\mF(\dyad{\psi})&=\sup\lset \Tr(W\psi) \sbar W\geq 0,\ \Tr(W\sigma)\leq 1,\ \sigma\in \mF \rset\\
 &\geq \sup\lset |\braket{w}{\psi}|^2 \sbar |\braket{w}{\phi}|\leq 1, \phi\in f \rset\\
 &=\bar{\xi}_f(\ket{\psi})
 \label{eq:robustness and extent bound}
\eal
where in the second line we restricted $W$ to the form $W=\dyad{w}$ for some unnormalized vector $\ket{w}$, and the third line is because of \eqref{eq:extent closure dual}.
This, together with \eqref{eq:extent closure equal robustness}, ensures that the optimal witness $\tilde W$ in \eqref{eq:assumption optimum achieved} takes the form $\tilde W=\dyad{\tilde w}$ for some unnormalized vector $\ket{\tilde w}$ satisfying $|\braket{\tilde w}{\phi}|\leq 1$ for every $\ket{\phi}\in f$.
Therefore, we get 
\bal
 \bar{\xi}_f(\ket{\psi}) & = \left|\sum_i \tilde c_i \braket{\tilde w}{\tilde\phi_i}\right|^2\\
 & \leq \left(\sum_i |\tilde c_i| \left|\braket{\tilde w}{\tilde\phi_i}\right|\right)^2\\
 & \leq \left(\sum_i |\tilde c_i|\right)^2\\
 & = \xi_f(\ket{\psi})
\eal
where in the first line we wrote $\ket{\psi}=\sum_i \tilde c_i \ket{\tilde \phi_i}$ and used that \eqref{eq:extent closure dual} is achieved with $\ket{\tilde w}$ due to \eqref{eq:robustness and extent bound}, the second line is due to the triangle inequality, and the third line is because $|\braket{\tilde w}{\phi}|\leq 1$, $\forall \ket{\phi}\in f$.
Then, \eqref{eq:assumption optimum achieved} implies that these all coincide, and in particular, 
\bal
 \braket{\tilde \omega}{\tilde\phi_i} = \frac{\tilde c_i}{|\tilde c_i|},\ \forall \tilde c_i \neq 0.
\eal
This gives  
\bal
\left|\Tr(\tilde W \dyad{\tilde\phi_i})\right| = 1,\ \forall i
\eal
concluding the proof.

\section{Numerical evaluation of maximal fidelity between Gaussian states and two Fock states}
\label{ap:numerics}
We want to compute the maximal overlap between a two-mode Gaussian state and two Fock states $\ket{1}\otimes \ket{1}$.
We numerically find an approximate solution of an optimization problem of maximizing the fidelity by parameterizing the unitaries in $\ket{G'}=U(\phi,\xi) S(r_1 e^{i\theta_1})\otimes S(r_2 e^{i\theta_2}) D(\alpha_1)\otimes D(\alpha_2)\ket{0}\otimes \ket{0}$, where $U$ is a passive symplectic unitary.
In this optimization, we numerically optimize the overlap between $\ket{G'}$ and $\ket{1}\otimes \ket{1}$.
We used the parameterization used in Ref.~\cite{PRXQuantum.2.020333}.
The parameters that we numerically found are
\begin{align}
    \alpha_1&=0,\\
    \alpha_2&=0,\\
    r_1&=0.8814,\\
    \theta_1&=0.609,\\
    r_2&=0.8814,\\
    \theta_2&=1.107,\\
    \phi&=-1.322,\\
    \xi&=1.571.
\end{align}
This yields
\begin{equation}
    \abs{\bra{1\otimes 1}\ket{G^\prime}}^2=0.25=\frac{1}{4},
\end{equation}
which coincides with the result reported in Ref.~\cite{Chabaud2025erratum}.


\bibliographystyle{quantum}
\bibliography{myref}

\end{document}